\documentclass{article}
\usepackage[utf8]{inputenc}
\usepackage[OT4]{fontenc}
\usepackage{graphicx} % Required for inserting images
\usepackage[left=2.5cm, right=2.50cm, top=3.20cm, bottom=3.20cm]{geometry}
\usepackage{amsfonts}
\usepackage{verbatim}
\usepackage{eucal}
\usepackage{amsmath}
\usepackage{amsthm}
\usepackage{mathrsfs}
\usepackage[unicode]{hyperref}
\usepackage{srcltx}
\newcommand{\arsinh}{\mathop {\rm arsinh}\nolimits }

\newcommand{\Vol}{\mathop {\rm Vol}\nolimits }
\newcommand{\sgn}{\mathop {\rm sgn}\nolimits }
\newcommand{\ddiv}{\mathop {\rm div}\nolimits }

\def\lin{\mathop{\rm span}}
\newcommand*{\Scale}[2][4]{\scalebox{#1}{$#2$}}%
\hypersetup{colorlinks=false}

\makeatletter
\newtheorem{df}{Definition}[subsection]
\newtheorem{thm}{Theorem}[subsection]

\makeatother
\title{Quasi-local mass on rigid or round spheres in Kerr spacetime}
\author{Jacek Jezierski, Tomasz Smo\l ka\thanks{Corresponding author:t.smolka@uw.edu.pl}\\
	Department of Mathematical Methods in Physics\\
	Faculty of Physics, University of Warsaw \\
	Pasteura 5, 02-093, Warsaw, Poland}
\date{August 26, 2024}
\begin{document}
	\numberwithin{equation}{section}
	\maketitle
	\begin{abstract}
		\noindent
		We introduce and analyze quasi-local mass using Hamiltonian methods. It is based on multipole decomposition for surfaces that are topological spheres. Based on the above model, tests were performed for Kerr spacetime for two arbitrary choices of surfaces: rigid spheres and round spheres.
	\end{abstract}
	\section{Introduction}
	Quasi-local mass is an essential concept in General Relativity that aims to quantify the mass-energy content of a spatially bounded region in curved spacetime. Unlike global concepts like ADM mass, which apply to asymptotically flat spacetimes at infinity, quasi-local mass should be defined for finite regions and is particularly useful in strongly gravitating systems with relatively large spacetime curvature. The quasi-local mass is designed to capture the energy due to matter and gravitational fields within a specified region, usually bounded by a closed two-surface. Expected properties of quasi-local mass are described in appendix \ref{ssec:QuasiMassProper}.\\
	In 1982, Penrose \cite{Penrose_list} presented a list of significant problems that remain to be solved. The most general was to ``find a suitable definition of quasi--local energy--momentum (mass)''. Over forty years have passed without an entirely satisfactory formulation of quasi--local mass. Construction of quasi-local mass is highly non-trivial because, in General Relativity, energy is not localizable due to the equivalence principle and the complicated interplay between matter and geometry in a manner consistent with the theory's non-linear, dynamical nature. This means that ``energy density'' for gravitational interactions can not exist. Hence, the understanding of energy is limited to the total mass of an isolated gravitational system. Despite such a difficult task, the potential benefits are still worth the effort. The concept of quasi-local mass is not just a theoretical curiosity. However, it has practical implications, for example, in studying black hole horizons, cosmological models, and the dynamics of gravitational systems. For details, see \cite{Szabados2009quasi}. 
	
	The ongoing research in this area continues to refine the definitions and applications of quasi-local mass, making it a central concept in understanding the gravitational interaction at astrophysical and cosmological scales. Moreover, the quasi-local mass plays a pivotal role in the study of isolated systems, such as black holes\footnote{For example, the recent analysis by Dunajski and Tod uses the isometric embedding of the spatial horizon of fast rotating Kerr black hole in a hyperbolic space to compute the Kijowski–Liu–Yau quasi-local mass of the horizon for any value of the spin parameter\cite{DunajskiTod2021}.} and stars, where it can be used to define the mass of the system in a way that accounts for both the local matter distribution and the geometry of the surrounding spacetime. In particular, the quasi-local mass also has implications for the study of gravitational radiation\footnote{ Lately, Chen, Wang, and Yau \cite{ChenWangYau2016} studied gravitational radiation by evaluating the Wang–Yau quasi-local mass of surfaces of fixed size at the infinity of both axial and polar perturbations of the Schwarzschild spacetime.}, as it allows for quantifying energy-momentum carried away by gravitational waves in a localized region of spacetime. This is particularly important for understanding the energy balance in processes such as binary black hole mergers, where traditional global mass concepts are inadequate. Furthermore, the quasi-local mass is instrumental in the thermodynamical description of spacetime, especially in black hole thermodynamics, which relates to the surface gravity and horizon area, providing insights into the fundamental laws governing black hole dynamics. Quasi--local mass is also especially useful in experiments because all experimental data comes from a finitely extended region. Moreover, all models describe a finite part of spacetime. Considering this, it should be no wonder that the issue of quasi-local mass is one of the most investigated topics in Astrophysics.\\
	There have been many attempts to define quasi-local mass. In the spacetimes with symmetries, we can formulate Komar quasi--local mass, which involves the use of the existence of a timelike Killing vector field and the associated closed 3-forms, which, due to the triviality of the third de Rham cohomology group in $\mathbb{M}^{4}$, are exact. This exactness allows for defining a 2-form associated with the Killing vector, which plays a significant role in expressing quasi-local conserved quantities. Arthur Komar formulated this quantity \cite{Komar_mass}  in 1959.\\
	In the case of spacetime with spherical symmetry, the construction of quasi--local quantities is usually based on a foliation of topological two-spheres, which are transitivity surfaces of the rotation group. Such spheres are essential when discussing quasi-local quantities such as the Misner-Sharp energy, which measures the energy contained within a ``round'' sphere. The Misner-Sharp energy is defined in terms of the Riemann curvature tensor components and the metric functions, and it provides a way to quantify the gravitational energy within a spherically symmetric spacetime. Initially, the construction discussed above was obtained by a heuristic comparison with the Schwarzschild metric (see \cite{Misner1964relativistic}, \cite{Hernandez1966observer} and \cite{Cahill1970spherical}).\\
	Yet another attempt to quasi--local energy has been done by Stephen Hawking \cite{Hawking_mass} in 1968. A result of Geroch \cite{Geroch_mon_inv_mean_curv} implies that Hawking--Geroch mass has an important monotone property along the inverse mean curvature flow. Two null, future-oriented vectors exist for each point of spacetime, defined upon a positive scale factor. Applying proper normalization, the Hawking--Geroch mass is defined as the integral over a two-dimensional Riemannian sphere isometrically embedded in a four-dimensional Lorentzian manifold from the product of such expansions\footnote{It can be described as follows: Let $\mathcal{S}$ be a two-dimensional Riemannian sphere isometrically embedded in a four-dimensional Lorentzian manifold. At each point $p \in \mathcal{S}$  We will denote them ${l}_{+}$ and ${l}_{-}$. The following demand can rig half of that freedom
		\begin{equation}
			g_{\mu \nu} {l}_{+}^{\mu} {l}_{-}^{\nu}=-2 \label{HG_norm}
		\end{equation}
		The expansion ${\theta}_{+}$ is defined by the divergences of the null hypersurfaces emanating from $\mathcal{S}$ tangentially to ${l}_{+}$
		\begin{equation}
			{\theta}_{+}=-\frac{1}{2}\tilde{\tilde{g}}^{AB}\pounds_{{l}_{+}}(g_{AB})=\nabla_{\mu}{l}_{+}^{\mu}
		\end{equation}
		and similarly for ${\theta}_{-}$.}.
	The definition of Hawking--Geroch mass plays an important role in the Geroch--Huisken--Ilmanen proof of the Penrose inequality (see \cite{Geroch_mon_inv_mean_curv} and \cite{Huisken2001inverse}). The Hawking--Geroch mass satisfies the monotonicity condition on asymptotically null or hyperbolic surfaces (see \cite{Jezierski1987_energy_positivity1}, \cite{Jezierski1987_energy_positivity2}, \cite{Kijowski1986_energy_positivity} and \cite{Sauter_mon_con_null_hyp}). However, it turns out that the Hawking--Geroch mass approaches the total mass only along a surface foliation that converges to a sphere of constant curvature after rescaling (see \cite{Neves_HG_mass_conv_spher_const_curv}). Particularly in the context of Hawking energy, the Gauss-Bonnet theorem and the Geroch–Held–Penrose formalism \cite{geroch1973space}  are usually employed to express quasi-local energy.\\
	Hamiltonian methods are also a powerful tool to construct quasi-local objects. Kijowski's mass \cite{Kijowski_var_form} is an important result in this context. It is based on the symplectic framework designed by Kijowski and Tulczyjew \cite{Kijowski_symp_framework}. The mass construction is based on external geometry encoded in the initial data set and the Weyl embedding theorem \cite{Nirenberg1953weyl,Pogorelov1951izgibanie}. An unexpected result of Liu and Yau \cite{Liu2003positivity} provides non-negativity of Kijowski mass. {\'O} Murchadha, Szabados, and Tod \cite{Murchadha2004comment}  show that the mass is nonzero for some surfaces in Minkowski spacetime, showing that the normalization is not optimal in general. This issue has been mentioned in \cite{Wang2006generalization}, \cite{Wang2009isometric}, the reader is referred to those works as well as \cite{Chen2011evaluating}, \cite{Wang2010limit} for further information and results.\\
	The Nester--Witten 2-form is another tool used to construct quasi--local mass, particularly in the Ludvigsen--Vickers\cite{ludvigsen1983momentum} and Dougan--Mason\cite{dougan1991quasilocal} constructions, which utilize spinor fields to define quasi--local masses.\\
	In 1958, Trautman \cite{Trautman1958} (see also \cite{Trautman2002king}) proposed a notion of energy suitable for asymptotically Minkowskian radiating gravitational fields and proved that it is monotonically decreasing. This idea has been further studied by Bondi et al. \cite{Bondi1962gravitational} and Sachs \cite{Sachs1962gravitational}. It has led to several other definitions of mass in the radiation regime. The Trautman--Bondi definition of mass is based on the existence of a particular coordinate system. The Trautman--Bondi mass is monotonically decreasing. It is a consequence of the famous Bondi mass-loss formula \cite{Bondi1962gravitational}.\\
	Passing to the more recent results, it is worth mentioning the Liu-Yau quasi-local mass \cite{Liu2003positivity}. The mass is based on geometrical construction, which involves the uniqueness of embedding a two-sphere into a flat three-dimensional space. This construction is gauge-invariant and does not rely on asymptotic flatness conditions, making it applicable in various physical scenarios. One of the key properties of the Liu-Yau quasi-local mass is its positivity, which has been proven under certain conditions \cite{Liu2003positivity}. This positive mass theorem for the Liu-Yau mass is significant as it aligns with the expectation that the energy content of a physical system should not be negative. This result generalizes Kijowski's ``free energy'' \cite{kijowski2002consistent}.\\
	Jezierski, Kijowski, and {\L}{\k{e}}ski \cite{Leski2013_Rigid_Spheres} proposed investigating the quasi-local mass of Kijowski free energy on foliations of topological spheres that imitate standard spheres in flat space. This approach is discussed in detail in later parts of the paper.\\
	Yet another definition of quasi-local mass is that proposed by Wang and Yau \cite{Wang2009isometric}. This definition is grounded in the Hamiltonian framework. It is based on the 'renormalized' form of the 'natural' Hamiltonian, which is a significant advancement in understanding the localized energy in gravitational systems. The quasi-local mass, according to Wang and Yau, is expressed through a reference isometric embedding into the Minkowski space, which allows for the comparison of the physical surface's geometry with that of a flat spacetime.\\
	However, there have been many attempts to give an appropriate definition of quasi-local mass during the last decades. The author recommends reading \cite{Szabados2009quasi}, \cite{Chrusciel2003hamiltonian_book}, and \cite{Chrusciel_lecture_notes} for a more detailed review.\\
	
	\subsection{Leading idea of the proposed research}

	The proposed analysis generalizes the Hamiltonian framework from \cite{Chr_Jez_Kij_Hamiltonian_Kerr_de_Sitter}. The approach describes Hamiltonian dynamics, including analysis of boundary terms (corner conditions). This leads to the proposition of quasi-local quantities, like mass (energy) or angular momentum, on the particular choice of foliations. In this context, the results by Jezierski, Kijowski, and {\L}{\k{e}}ski \cite{Leski2013_Rigid_Spheres} provide the necessary proof of the existence of such foliations. They proposed investigating the quasi-local mass of Kijowski free energy on foliations of topological spheres that imitate standard spheres in flat space. Their intrinsic geometry defines these spheres, called round spheres\footnote{Different round spheres -- Round spheres are defined within spherically symmetric spacetimes. These are two-dimensional surfaces that are invariant under the action of the rotation group $SO(3)$, which means they are surfaces of constant curvature that are characterized by their symmetry under rotations. A round sphere can be identified as a transitivity surface for this group in a spherically symmetric spacetime. This implies that the group's action can transitively map any point on the sphere to any other point on the same sphere.\\
		The concept of round spheres is fundamental when discussing quasi-local quantities such as the Misner-Sharp energy, which measures the energy contained within a round sphere. The Misner-Sharp energy is defined in terms of the Riemann curvature tensor components and the metric functions, and it provides a way to quantify the gravitational energy within a spherically symmetric spacetime. Initially, the mass discussed above was obtained by a heuristic comparison with the Schwarzschild metric (see \cite{Misner1964relativistic}, \cite{Hernandez1966observer} and \cite{Cahill1970spherical}).},  by constant two-dimensional scalar curvature. Furthermore, a topological sphere with a constant Ricci scalar on the whole sphere is isotropic and homogeneous, meaning it has the same properties at every point and direction. This property is adapted to physical applications, particularly in cosmology, where the cosmological principle assumes that the universe is homogeneous and isotropic on large scales.
	
	Another proposition involves simple exterior geometry, which implies a constant mean curvature. Such topological spheres are called rigid spheres. This paper aims to examine how such defined quasi-local masses are a good description of the mass in asymptotically Kerr spacetimes.
	
	\subsection{Notation and Conventions}
	
	We wish to define a Hamiltonian dynamical system on a set of Lorentzian metrics with signature $(-,+,+,+)$ on a four-dimensional manifold $\mathcal{M}$, assuming the existence of a three-dimensional spacelike surface $\Sigma \subset \mathcal{M} $ on which the metric approaches a Kerr metric, see (\ref{kerr.metric}), as one recedes to infinity along an end of $\Sigma$.\\
	In order to present our results, we introduce some notation analogical to \cite{Chr_Jez_Kij_Hamiltonian_Kerr_de_Sitter}. $\mathcal{M}$ is an four-dimensional spacetime equipped with a Lorentzian metric $g_{\mu \nu}$. Consider a spacetime domain $\Omega$ with a smooth timelike boundary such that $\mathcal{V}:=\Omega \cap \Sigma$ is compact. Let $x^{\mu}=(x^{0},x^{k})$ be coordinates on $\Omega$ such that $x^{0}$ is a time coordinate constant on $\Sigma$ and $x^{k}$, where $k$ runs $k,l..z$ except $r$, are local coordinates on $\Sigma$. By $x^{a}=(x^{0},x^{A})$, where $a$ runs $a,b...j$, we mean coordinates on a world tube $\partial \Omega$ with coordinates $x^{A}$ on $\partial \mathcal{V}$. We distinguish $x^{r} \in \{x^{k} \}$ which is constant on $\partial \mathcal{V}$.   In the appendix \ref{sec:RigidFourDimAppendix} only, we introduce the tilded Greek indices to denote normal directions to a topological sphere. The covariant derivative associated with the four-dimensional metric will be denoted by $\nabla$ or just by ";". We will denote by $T_{...(\mu\nu)...}$ the symmetric part and by $T_{...[\mu\nu]...}$ the antisymmetric part of tensor $T_{...\mu\nu...}$ concerning indices $\mu$ and $\nu$ (analogous symbols will be used for more indices).\\
	In the paper, the multipole decomposition is introduced on two particular choices of topological spheres. For a given scalar field, say $\Phi$, we distinguish monopole and dipole parts of $\Phi$, denoted by $\Phi_m$ and $\Phi_d$ respectively. Additionally, the higher multipoles are distinguished. More precisely, we define
	\begin{equation}
		\Phi_{w}=\Phi-\Phi_{m}-\Phi_{d} \, .
	\end{equation}
	The multipole decomposition can be obtained with the help of the equilibrated coordinate system. See appendix \ref{multipole_expansion} for details.
	It is convenient to introduce some notation and present basic identities for the analysis of variational formulas. By $S_{ab}$ we denote the extrinsic curvature tensor of $\partial \Omega$ embedded in $\mathcal{M}$
	\begin{equation}
		S_{ab}=-\frac{ \Gamma^{r}_{ab}}{\sqrt{g^{rr}}}=-\frac{A^{r}_{ab}}{g^{rr}} \, , \label{extr_curv}
	\end{equation}
	where $\Gamma^{n}_{pq}$ denotes a Christoffel symbol and $A^{r}_{ab}$ is defined in (\ref{def_A}). $Q_{ab}$ is the ``Arnowitt--Deser--Misner (ADM)'' counterpart of (\ref{extr_curv})
	\begin{equation}
		Q^{ab}:=\sqrt{|\det g_{ef}|}(S\tilde{g}^{ab}-S^{a b}) \, , \label{ADM_cpart}
	\end{equation}
	where $\tilde{g}^{ab}$ is the three-dimensional inverse with respect to the induced metric $g_{ab}$ on the world tube $\partial \Omega$ (see Appendix \ref{time_decomposition}).\\
	The objects on $\partial \mathcal{V}$ will be defined. We start with the decomposition of the time coordinate from the three-dimensional metric on $\partial \Omega$ (see appendix \ref{time_decomposition}).
	One can define the following two-dimensional objects on $\partial \mathcal{V}$:\\
	a scalar density
	\begin{equation}
		\textbf{Q}:=\nu Q^{00} \label{bold_Q} \, ,
	\end{equation}
	and a covector density
	\begin{equation}
		\textbf{Q}_{A}:={Q^{0}}_{A}=Q^{0b}g_{bA} \, .
	\end{equation}
	It is further useful to define the perpendicular component of $Q^{ab}$
	\begin{equation}
		\mathbf{Q}^{AB}:=\frac{1}{\nu}Q_{CD} \tilde{\tilde{g}}^{CA} \tilde{\tilde{g}}^{DB} \, , \label{perp_Q}
	\end{equation}
	where $ \tilde{\tilde{g}}^{CA}$ is an inverse of the metric on $\partial \mathcal{V}$. We also have
	\begin{equation}
		S^{00}=S_{ab}\tilde{g}^{0 a} \tilde{g}^{0b}=\frac{1}{\nu^{4}} \left(S_{00}-2 S_{0A} \nu^{A}+S_{AB} \nu^{A} \nu^{B}\right) \, ,
	\end{equation}
	with the trace $S$ of $S_{ab}$ being equal to
	\begin{eqnarray}
		S&=&S_{ab} \tilde{g}^{ab}=S_{00}\tilde{g}^{00}+2 S_{0A} \tilde{g}^{0A}+S_{AB}\tilde{g}^{AB} \nonumber \\
		&=&-\frac{1}{\nu^2} S_{00}+2 S_{0A}\frac{\nu^A}{\nu^2}+S_{AB}\left( \tilde{\tilde{g}}^{AB}-\frac{\nu^A \nu^{B}}{\nu^2} \right) \nonumber\\
		&=&-\nu S^{00}+S_{AB} \tilde{\tilde{g}}^{AB}=a-k \, , \label{S}
	\end{eqnarray}
	where $k$ is a trace of the two-dimensional part of the extrinsic curvature $S_{AB}$. By
	\begin{equation}
		a:=-v^2 S^{00} \, ,
	\end{equation}
	we denote the curvature ``acceleration'' of the world lines which are geodesic within $\partial \Omega$ and orthogonal to $\partial \Omega \cap \Sigma$. The following relations hold
	\begin{eqnarray}
		\textbf{Q}&=&\nu^2 \lambda \left(S \tilde{g}^{00}-S^{00}\right) \nonumber \\
		&=&\lambda \left(-S-\nu^2 S^{00}\right)=\lambda k \, , \label{Q_lambda_k}
	\end{eqnarray}
	and for (\ref{perp_Q})
	\begin{eqnarray}
		\mathbf{Q}^{AB}g_{AB}&=&\frac{1}{\nu}Q_{CD}\tilde{\tilde{g}}^{CD} \nonumber\\
		&=& \lambda \left(S g_{AB}-S_{AB}\right) \tilde{\tilde{g}}^{AB} \nonumber \\
		&=& \lambda \left(2 a-k\right) \, , \label{bold_Q_2dim}
	\end{eqnarray}
	where
	\begin{equation}
		\lambda:=\sqrt{\det g_{AB}} \, ,
	\end{equation}
	is the two-dimensional area element on $\partial \mathcal{V}$. Let $\alpha$ be the hyperbolic angle between $\partial \Omega$ and $ \partial \mathcal{V}$. In the adapted coordinates presented above\footnote{Let us recall that $x^{r}$ is the coordinate which is constant on $\partial \Omega$ and $x^{0}$ is a time coordinate.}
	\begin{equation}
		\alpha:=\arsinh \left(\frac{g^{0r}}{\sqrt{|g^{00}g^{rr}|}}\right) \, . \label{alpha}
	\end{equation}
	The above definition demands non-vanishing $\nu$. The hypothesis that $\nu$ has no zeros is needed to derivate the elegant formula (\ref{kij_var_form}). However, we stress that this assumption can be neglected.\\
	
	\section{Examination of quasi-local mass in Kerr spacetime}
	
	\subsection{Motivation and brief description of investigated issue}
	\label{sec:Sph_motivation}

	Typically, the $S^2$-spheres used for the quasi-local purposes come from a specific choice of coordinates $t$ and $r$, which play the role of a gauge. The most used gauge conditions in the scientific literature are based on three-dimensional elliptic problems (see, e.g.~``traceless-transversal'' condition advocated by York analyzed in \cite{York1974covariant}). Jezierski has obtained important results (see \cite{Jezierski1994stability}) who used a parabolic gauge condition to prove the stability of the Reissner--Nordstr\"{o}m solution, together with a version of Penrose's inequality. These approaches exhibit an obvious drawback consisting in the fact that we do not control the intrinsic properties of the surfaces $\{r=\mathrm{const.}\}$ constructed this way. Huang partially resolved this problem (see \cite{Huang2010_mean_spheres}). Their fibers $\{r=\ $const.$\}$ are selected by two-dimensional elliptic equation: $k=\mathrm{const.}$, where $k$ is the mean extrinsic curvature. Unfortunately, the above condition is not stable, especially concerning small perturbations of the geometry. This paper presents our results, in which we test two surfaces that do not exhibit the above drawback. Those spheres are discussed in detail in the next paragraphs.
	\subsubsection{A particular choice of studied surfaces: Rigid Spheres and Round Spheres}
	\label{ssec: Round_Rigid_definition}

	\paragraph{Round Spheres} are two-dimensional, closed surfaces whose intrinsic geometry is assumed to be controlled:
	\begin{df}
		A Round Sphere is a topological two-dimensional sphere whose inner geometry is round. It means that the curvature scalar is constant
		\begin{equation}
			R^{(2)}(\left. \vphantom{Q} g \right|_{\mathcal{S}})=\mathrm{const.}
		\end{equation}
		or, in other words, it is a constant curvature space.
		\label{def:round_sphere}
	\end{df}
	In this case, a coordinate system exists in which the induced, two-dimensional metric is, up to a multiplicative constant, a standard spherical metric $\sigma_{AB}$. It means that the multipole expansion defined in appendix \ref{multipole_expansion} is well-defined.
	\paragraph{Rigid Spheres} are an improved idea of Huang proposed by Gittel, Jezierski, Kijowski, and Łęski in \cite{Leski2013_Rigid_Spheres}. Originally, the authors defined the multipole expansion based on Equilibrated coordinates\footnote{These methods are briefly described in appendix \ref{multipole_expansion}.}. The condition on the wave part of mean extrinsic curvature defines Rigid Spheres:
	\begin{df}
		Let $\Sigma$ be a Riemannian three-manifold and let $\mathcal{S} \subset \Sigma$ be a two-dimensional, topological sphere. We say that $\mathcal{S}$ is a \emph{rigid sphere} if its mean extrinsic curvature $k$ satisfies the following equation:
		\begin{equation}\label{rig}
			k_{w} = 0 \, ,
		\end{equation}
		which means that the external curvature vanishes modulo mono-dipole part.
	\end{df}
	
	The condition (\ref{rig}) is also a two-dimensional elliptic equation, but it is weaker than ``$k=\mathrm{const.}$''. The more detailed analysis of Rigid Spheres in four-dimensional Lorentzian spacetime is described in appendix \ref{sec:RigidFourDimAppendix}.
	
	In the context of the above--stated definitions, a natural question arises: When are the Round Spheres and the Rigid Spheres the same class of surfaces? Consider a three-dimensional Riemannian manifold with a metric $(\Sigma, g_{kl})$. We distinguish a two-dimensional surface $\mathcal{S}$. It is convenient to split the three-dimensional Riemann tensor into a two-dimensional Riemann tensor induced on $\mathcal{S}$ and an external curvature of $\mathcal{S}$ embedded in $\Sigma$
	\begin{equation}
		{}^{(3)}R_{ABCD}={}^{(2)}R_{ABCD}-k_{AC} k_{BD}+k_{AD}k_{BC} \, . \label{Gauss_id}
	\end{equation}
	We also decompose the external curvature tensor into its traceless part and the rest
	\begin{equation}
		k_{AB}=\check{k}_{AB}+ \frac12 k g_{AB} \, .
	\end{equation}
	Contracting the equation (\ref{Gauss_id}) with the inverse of two-dimensional metric, we obtain
	\begin{equation}
		{}^{(3)}R_{ABCD} \tilde{\tilde{g}}^{CA} \tilde{\tilde{g}}^{DB}={}^{(2)}R- \frac12 k^{2}+\check{k}_{AB} \check{k}^{BA} \, . \label{Gauss_eq_con}
	\end{equation}
	Consider the following operator
	\begin{equation}
		U={}^{(3)}R_{ABCD} \tilde{\tilde{g}}^{CA} \tilde{\tilde{g}}^{DB}-\check{k}_{AB} \check{k}^{BA}={}^{(2)}R- \frac12 k^{2}=\mathrm{const.} \label{Gauss_con}
	\end{equation}
	If we demand that $\mathcal{S}$ fulfills the definition of a Round Sphere and the definition of a Rigid Sphere simultaneously for each point $p \in \mathcal{S}$, then
	\begin{equation}
		U(p)=\mathrm{const.} \label{Round_Rigid_condition}
	\end{equation}
	{The above condition is naturally fulfilled for spacetimes with spherical symmetry. It can hold for spacetimes for which the symmetry is perturbed. We do not want to analyze in detail the existence of such surfaces. Note that a class of embeddings exists for which \eqref{Round_Rigid_condition} is obeyed.}
	
	The main idea of this section is to investigate a quasi-local mass on two-dimensional surfaces that mimic the case of flat Minkowski space. In other words, our surfaces create a family of standard spheres embedded in all possible flat subspaces $\Sigma \subset \mathcal{M}$. Below, we analyze two particular classes of such surfaces: Rigid Spheres and Round Spheres. The reach geometrical structure of the selected surfaces gives a potentially significant chance to set specialized gauge conditions. The variational formula in general form has the following form:
	\begin{eqnarray}
		\hspace{- 1 cm}
		\int_{\mathcal{V}} \left[\partial_{0} P^{kl} \delta g_{kl}-\partial_{0} g_{kl} \delta P^{kl}\right] \mathrm{d}\Sigma_{0}& & \nonumber \\
		+2 \int_{\partial \mathcal{V}} \left[\partial_{0}\lambda \delta \alpha-\partial_{0}\alpha \delta \lambda \right] \mathrm{d} \mathcal{S}_{0r}&=&-\int_{\partial \mathcal{V}} \left[2 \nu \delta \textbf{Q}-2 \nu^{A} \delta \textbf{Q}_{A} + \nu \textbf{Q}^{AB} \delta g_{AB} \right] \mathrm{d} \mathcal{S}_{0r} \nonumber\\
		&=& - 16 \pi  \int_{\partial \mathcal{V}} \delta \widetilde{\mathcal{H}}_{\partial \mathcal{V}}\mathrm{d} \mathcal{S}_{0r}-\int_{\partial \mathcal{V}} g_{ab}\delta Q^{ab}\mathrm{d} \mathcal{S}_{0r} \, ,
		\hspace{0.2 cm} \label{almost_Hamiltonian}
	\end{eqnarray}
	where $\widetilde{\mathcal{H}}_{\partial \mathcal{V}}$ pretends to be a quasi-local Hamiltonian generating function. The proof of the formula is presented in appendix \ref{variational_Kijowski}. The final form is given by  (\ref{kij_var_form}). The Hamiltonian dynamics for General Relativity can be represented as a boundary integral over a two-dimensional surface. It causes significant problems distinguishing the contribution to the Hamiltonian from the control parameters. For a system to be well-defined, we can impose conditions on two real degrees of freedom and four gauge fixing conditions on the boundary. If we do not set gauge conditions, the equation (\ref{almost_Hamiltonian}) leads to a Hamiltonian, giving unexpected results for Kerr spacetime. In particular, an unwanted correction related to angular momentum exists. On the other hand, when the ADM counterpart remains fixed ($\delta Q_{ab}=0$), an essential contribution to the mass is not included. The linearized theory of gravity has some hints\footnote{See the equations (\ref{real_degree_on_dV}) and (\ref{QA_w_gauge}) and the comments below them.} for us about Hamiltonian properties. Anderson performed similar considerations in \cite{anderson2008boundary}. In his paper, the spatial part of the metric is given as an initial value problem.
	
	\subsection{Gauge conditions for multipole expansion \label{sec: Gauge_conditions}}
	We aim to make a multipole decomposition (see \ref{multipole_expansion} for details) and set appropriate gauge conditions for multipole expansion.
	Making use of (\ref{Q_lambda_k})
	\begin{equation}
		2 \nu \delta \textbf{Q}=\lambda \frac{\nu}{k} \delta (k^2)+2 \nu k \delta(\lambda) \label{H_analyze}
	\end{equation}
	Each vector can be decomposed into transversal and longitudinal parts (see \ref{multipole_expansion} for notation). Using (\ref{H_analyze}), we obtain multipole decomposition of the crucial ingredient of (\ref{almost_Hamiltonian})
	\begin{eqnarray}
		\int_{\partial \mathcal{V}} \left[2 \nu \delta \textbf{Q}-2 \nu^{A} \delta \textbf{Q}_{A} \right] \mathrm{d} \mathcal{S}_{0r}&=& \int_{\partial \mathcal{V}}\left \{\lambda \left[\left(\frac{\nu}{k} \right)_{m}+\left(\frac{\nu}{k} \right)_{d}+\left(\frac{\nu}{k} \right)_{w}\right] \delta (k^2_{m}+k^2_{d}+k^2_{w}) \right. \nonumber \\
		& &-\left.2 \nu_{A} \delta \left({\! \hphantom{|}}_{\perp \! \!} \mathbf{Q}^{A}_{d}+{\! \hphantom{|}}_{|| \!} \mathbf{Q}^{A}_{d}+{\! \hphantom{|}}_{\perp \! \!} \mathbf{Q}^{A}_{w}+{\! \hphantom{|}}_{|| \!} \mathbf{Q}^{A}_{w}\right)  \right \} \mathrm{d} \mathcal{S}_{0r}
		\label{H_multipole}
	\end{eqnarray}
	Making use of the orthogonality of the multipoles
	\begin{eqnarray}
		\int_{\partial \mathcal{V}} \left[2 \nu \delta \textbf{Q}-2 \nu^{A} \delta \textbf{Q}_{A} \right] \mathrm{d} \mathcal{S}_{0r}= \int_{\partial \mathcal{V}}\left \{\lambda \left[\left(\frac{\nu}{k} \right)_{m} \delta (k^2_{m})+\left(\frac{\nu}{k} \right)_{d} \delta (k^2_{d})+\left(\frac{\nu}{k} \right)_{w} \delta (k^2_{w})\right] \right. \nonumber \\
		-\left.2 \left[\left({\! \hphantom{|}}_{\perp \!} \nu_{A} \right)_{ d}\delta({\! \hphantom{|}}_{\perp \! \!} \mathbf{Q}^{A}_{d})+ \left({\! \hphantom{|}}_{|| \!} \nu_{A} \right)_{ d} \delta ( {\! \hphantom{|}}_{|| \!} \mathbf{Q}^{A}_{d})+ \left({\! \hphantom{|}}_{\perp \!} \nu_{A} \right)_{ w} \delta ( {\! \hphantom{|}}_{\perp \! \!} \mathbf{Q}^{A}_{w})+ \left({\! \hphantom{|}}_{|| \!} \nu_{A} \right)_{ w} \delta ({\! \hphantom{|}}_{|| \!} \mathbf{Q}^{A}_{w})\right]  \right \} \mathrm{d} \mathcal{S}_{0r} \hspace{0.5 cm}	\label{H_multipole_ort}
	\end{eqnarray}
	For both classes of foliations used in the paper, we propose the following gauge
	\begin{enumerate}
		\item The dipole part of $k^{2}$ is related to the translation of a sphere. We set the center of a sphere in the origin on the boundary. It is equivalent to
		\begin{equation}
			\delta k^{2}_{d}|_{\partial \mathcal{V}}=0 \, . \label{translation_gauge}
		\end{equation}
		\item The wave part of $k^{2}$ is treated as a real degree of freedom. we require to set $k^{2}_{w}$ fixed on the boundary
		\begin{equation}
			\delta k^{2}_{w}|_{\partial \mathcal{V}} = 0 \, , \label{real_degree_on_dV}
		\end{equation}
		which can be understood as the Dirichlet initial conditions.
		\item The monopole part of $k^{2}$ is the main contribution to the mass. We do not impose conditions on this term and its perturbation.
	\end{enumerate}
	Additionally, we assume that the two-dimensional metric on the boundary $\mathcal{S}$ is fixed, which means
	\begin{equation}
		\delta g_{AB}|_{\partial \mathcal{V}}=0 \, . \label{2-D_metric_control}
	\end{equation}
	Some detailed comments about our Hamiltonian system are in order. In our model, we examine the energy within a topological sphere. Therefore, our region's boundary is either a round sphere or a rigid sphere, depending on the case. If we want to investigate the change in energy between spheres, we treat each system, whose region is a bulk surrounded by the sphere, as independent. Therefore, we solve constraint equations in volume with boundary conditions set on the sphere. We do not intend to solve field equations between two topological spheres.
	Additionally, the equations \eqref{Gauss_id}-\eqref{Round_Rigid_condition} and comments nearby lead to the conclusion that the surface to be both a rigid sphere and a round sphere simultaneously, exist only for a very restricted class of spacetimes. Therefore,  in the case of a round sphere, we impose a condition on the form of two-dimensional metrics. At the same time, the non-monopole part of the external curvature is fixed but arbitrary (its perturbation is equal to zero). When considering a foliation consisting of rigid spheres, the curvature has a specific form, while the two-dimensional metric is fixed but arbitrary (with zero perturbation). It allows us to minimize restrictions on the class of spacetimes that can be considered in our framework.
	
	Now we pass to impose conditions on  $2 \nu^{A} \delta \mathbf{Q}_{A}$
	\begin{itemize}
		\item[4.] The freedom related to the boost transformations can be fixed by
		\begin{equation}
			\left.\ddiv \mathbf{Q}^{A}\right|_{\partial \mathcal{V}}=0 \, \Leftrightarrow \, \left.\mathbf{Q}^{A} \right|_{\partial \mathcal{V}}={\! \hphantom{|}}_{\perp \! \!} \mathbf{Q}^{A} \, .
		\end{equation}
		\item[5.] The wave part of $\mathbf{Q}^{A}$ is a real degree of freedom. The Dirichlet conditions are set as
		\begin{equation}
			\left. \delta \mathbf{Q}^{A}_{w} \right|_{\partial \mathcal{V}}=0 \, . \label{QA_w_gauge}
		\end{equation}
		The idea of the gauges (\ref{real_degree_on_dV}) and (\ref{QA_w_gauge})  has its origins in the weak field theory. If we expand the metric to the quadratic term, we can construct gauge invariant quantities, called $\mathbf{x}$ and $\mathbf{y}$, which correspond respectively to $ k^{2}_{w}$ and $\mathbf{Q}^{A}_{w}$. In \cite{jezierski1990_linear_hamiltonian} and \cite[section 3.2]{2018smojez_hopf}, the local Hamiltonian for the weak gravitational field was presented. It naturally suggests the decomposition of $k$ and $\mathbf{Q}^{A}$ into multipoles.
	\end{itemize}
	The above gauge conditions (\ref{translation_gauge})-(\ref{QA_w_gauge}) define an one-parameter family of spheres. We do not specify how to compare the passage of time on the spheres with different radial coordinates. We introduce the following requirements:
	\begin{itemize}
		\item[6.] Standardization
		\begin{equation}
			-\left(\frac{\nu}{k}\right)_{m}=\sqrt{\frac{A}{16 \pi}} \label{standardization}
		\end{equation}
		where $A$ is the area of a sphere.
		
		Taking into account the conditions (\ref{translation_gauge})-(\ref{standardization}), the quasi-local Hamiltonian generating function (\ref{H_analyze}) has the form
		\begin{equation}
			\delta \mathbf{H}_{\partial \mathcal{V}}= \frac{1}{16 \pi} \int_{\partial {\mathcal{V}}} \left[\lambda \left(\frac{\nu}{k} \right)_{m} \delta (k^2_{m})- 2 \left({\! \hphantom{|}}_{\perp \!}\nu_{A} \right)_{ d} \delta {\! \hphantom{|}}_{\perp \! \!} \mathbf{Q}^{A}_{d}\right] \mathrm{d} \mathcal{S}_{0r} \, . \label{pre_generating_hamiltonian_formula}
		\end{equation}
		In the next step, we process the following term
		\begin{equation}
			- 2 \left({\! \hphantom{|}}_{\perp \!}\nu_{A} \right)_{ d} \delta {\! \hphantom{|}}_{\perp \! \!} \mathbf{Q}^{A}_{d}=- 2 \left({\! \hphantom{|}}_{\perp \!}\nu_{A} \right)_{ d} \delta \left( {\! \hphantom{|}}_{\perp \! \!} \mathbf{Q}^{A}_{d} - c \lambda {\! \hphantom{|}}_{\perp \!}\nu^{A} \right)-2 \left({\! \hphantom{|}}_{\perp \!}\nu_{A} \right)_{ d} \delta \left[ c \lambda \left({\! \hphantom{|}}_{\perp \!}\nu^{A} \right)_{ d} \right] \, . \label{c_introduce}
		\end{equation}
		\item[7.] We propose the gauge condition for the shift vector on boundary $\partial \mathcal{V}$
		\begin{equation}
			{\! \hphantom{|}}_{\perp \! \!} \mathbf{Q}^{A}_{d} = c \lambda \left({\! \hphantom{|}}_{\perp \!}\nu^{A} \right)_{ d}
		\end{equation}
		where $c$ is an arbitrary, scalar function depending on $r$ and $\theta$. The Legendre transformation is performed on (\ref{c_introduce})
		\begin{equation}
			- 2 \left({\! \hphantom{|}}_{\perp \!}\nu_{A} \right)_{ d} \delta {\! \hphantom{|}}_{\perp \! \!} \mathbf{Q}^{A}_{d}=-2 \delta \left[ \left({\! \hphantom{|}}_{\perp \!}\nu_{A} \right)_{ d} \left( c \lambda \left({\! \hphantom{|}}_{\perp \!}\nu^{A} \right)_{ d} \right) \right]+2 c \lambda \left({\! \hphantom{|}}_{\perp \!}\nu^{A} \right)_{ d} \delta \left({\! \hphantom{|}}_{\perp \!}\nu_{A} \right)_{ d} \label{c_to_hamiltonian}
		\end{equation}
	\end{itemize}
	Using(\ref{pre_generating_hamiltonian_formula}) and (\ref{c_to_hamiltonian}), the following Hamiltonian generating formula is received
	\begin{eqnarray}
		\delta \mathbf{H}_{\partial \mathcal{V}}&=& \frac{1}{16 \pi} \int_{\partial {\mathcal{V}}} \left[\lambda \left(\frac{\nu}{k} \right)_{m} \delta (k^2_{m})+2 c \lambda \left({\! \hphantom{|}}_{\perp \!}\nu^{A} \right)_{ d} \delta \left({\! \hphantom{|}}_{\perp \!}\nu_{A} \right)_{ d}\right] \mathrm{d} \mathcal{S}_{0r} \nonumber \\
		&=& \frac{1}{16 \pi} \int_{\partial {\mathcal{V}}} \left \{\lambda \left(\frac{\nu}{k} \right)_{m} \delta (k^2_{m})+c \lambda  \delta \left[ \left({\! \hphantom{|}}_{\perp \!}\nu^{A} \right)_{ d} \left({\! \hphantom{|}}_{\perp \!}\nu_{A} \right)_{ d} \right]\right \} \mathrm{d} \mathcal{S}_{0r} \label{generating_hamiltonian_formula}
	\end{eqnarray}
	It leads to the following Hamiltonian
	\begin{equation}
		\mathbf{H}_{\partial \mathcal{V}}
		=\frac{1}{16 \pi} \int_{\partial {\mathcal{V}}} \left \{\lambda \left(\frac{\nu}{k} \right)_{m} [(k^2)_{m}-(k_{0}^2)_{m}]+ \left({\! \hphantom{|}}_{\perp \!} \nu_{ A} \right)_{d} {\! \hphantom{|}}_{\perp \! \!} \mathbf{Q}^{A}_{d} \right \}\mathrm{d} \mathcal{S}_{0r} \label{Hamiltonian_multipoled_final}
	\end{equation}
	where $k_{0}$ is the mean curvature of the reference frame. It is a mean curvature of $\mathcal{S}$ embedded in three-dimensional Euclidean space.
	\section{Results - approach based on asymptotic series}
	
	In this section, we obtain Hamiltonian dynamics for two particular choices of surfaces: Round Spheres and Rigid Spheres. The properties mentioned below are valid for both models. The unique features of each surface class are presented in the following sections. {More precisely, we obtain an asymptotic series of explicit formulae for all objects involved in the Hamiltonian formula (\ref{H_multipole_ort}).}
	
	We start with the Kerr spacetime in Boyer--Lindquist coordinates. Locally, the Kerr solution to the vacuum Einstein equations is the metric $g_{\mu\nu}$, which in Boyer--Lindquist coordinates takes the form
	\begin{eqnarray}
		\label{kerr.metric}
		\nonumber
		g_{\mu\nu}\mathrm{d}x^\mu \mathrm{d}x^\nu &=& \rho^2\left(\frac{1}{\Delta} \mathrm{d}r^2+ \mathrm{d}\theta^2\right)
		+\frac{\sin^2\theta}{\rho^{2}} \left(a \mathrm{d}t-(r^2+a^2)\mathrm{d}\varphi\right)^2  \\
		&\quad& -  \frac{\Delta}{\rho^{2}}\left(\mathrm{d}t-a \sin^2 \theta \, \mathrm{d}\varphi\right)^2 %\;,
	\end{eqnarray}
	where
	\begin{eqnarray}
		\label{rho.kerr}
		\rho^2 &=& r^2+a^2\cos^2 \theta \vphantom{\frac11} %\;,
		\\
		\label{delta.kerr}
		\Delta  &=& (r^2+a^2) -2 m r %\;,
	\end{eqnarray}
	with $t \in \mathbb{R}$, $r \in \mathbb{R}$, and $\theta$, $\varphi$ being the standard coordinates parameterizing a
	two-dimensional sphere. We will avoid zeros of $|\rho|$ and $\Delta$ and ignore the coordinate singularities $\sin \theta=0$. This metric describes a rotating object of mass $m$ and angular momentum $J=ma$. The advantage of the above coordinates is that for $r$ much greater than $m$ and $a$, the metric  becomes asymptotically flat, i.e. $g \approx -\mathrm{d} t^2+\mathrm{d} r^2 +r^2 (\mathrm{d} \theta^2+\sin^2 \theta \mathrm{d} \phi^2)$.
	
	The selected surface is embedded in the spatial slice $\{x^{0}=\mathrm{const.}\}$ of Kerr spacetime. Boyer--Lindquist coordinates are the initial coordinates. Both classes of spheres have axial symmetry.
	\subsection{Boundary conditions for the surface equations}
	In both cases, the definition of the surface leads to a second-order partial differential equation. For the Rigid Sphere, it is a two-dimensional elliptic equation. In the case of Round Sphere, the equation is highly non-linear. The properties mentioned below are valid for both models. The solution has the form
	\begin{equation}
		R=f(r,\theta)
	\end{equation}
	The explicit, analytical form of the solution for both Rigid and Round surfaces is unknown. The following form of the asymptotic series is proposed
	\begin{equation}
		R=\sum_{j=1}^{-\infty} c_{j}(\theta) r^j=c_{1}(\theta) r+c_{0}(\theta)+\frac{c_{-1}(\theta)}{r}+...
	\end{equation}
	We set the conditions for the series:
	\begin{enumerate}
		\item $c_{1}=1$\\
		It can be treated like a boundary condition in spatial infinity. We demand that the surface approaches a standard sphere in the asymptotically flat regime.
		\item The solution has the symmetry $c_{i}(\theta)=c_{i}(\pi-\theta)$.
		\item The functions $c_{i}(\theta)$ have not singularities for $\theta \in [0,\pi]$
	\end{enumerate}
	We have not yet obtained a complete proof of the convergence of the series. However, it turns out the above conditions define unambiguously the functions $c_{0}(\theta),c_{-1}(\theta),...,c_{-6}(\theta)$ (up to calculated precision in (\ref{R_series}) and (\ref{R_rigid_series})). We believe that it holds for the whole solution in the series form. The convergence issue for the solution will be investigated in a separate paper.
	\subsection{Results for Round Spheres}
	\label{ssec:Results_Round}
	
	Let $\mathcal{S}$ be a two-dimensional surface embedded in spatial slice $\Sigma=\{x^{0}=\mathrm{const.}\}$. In our case, $\mathcal{S}$  plays the role of $\partial \mathcal{V}$ defined at the beginning of section \ref{variational_Kijowski}. We describe $\mathcal{S}$ by the following equation
	\begin{equation}
		R=f(r,\theta)=\mathrm{const.} \label{def_R}
	\end{equation}
	where $f(r,\theta)$ is smooth. On the other hand, we obtain a foliation of spheres where $R$ parameterizes the leaves. It is convenient to perform a two-dimensional transformation of coordinates
	\begin{equation}
		(r,\theta) \to (R, \Theta) \label{Coord_transform}
	\end{equation}
	where $R \in \mathbb{R}_{+}$ is implicitly given by the equation (\ref{def_R}).  $\Theta \in (0,\pi)$ is defined unambiguously by the following equations
	\begin{eqnarray}
		\sin \Theta&=&\frac{\sqrt{(r^2+a^2)\rho^2 +2mra^2 \sin^2 \theta}}{R \rho} \sin \theta \nonumber\\ & & \label{sin_Th} \\
		\cos \Theta&=&\frac{\sqrt{[R^{2}-(r^{2}+a^{2})\sin^{2} \theta] \rho^2 - 2 m r a^2 \sin^4 \theta}}{R \rho} \nonumber
	\end{eqnarray}
	where $\rho$ is defined by (\ref{rho.kerr}).
	The equations (\ref{sin_Th}) fulfill basic trigonometric properties, i.e., $\sin^{2} \Theta+\cos^{2} \Theta=1$. Moreover, $\Theta$ is constructed in a way that fulfills the condition
	\begin{equation}
		g_{\varphi \varphi}=R^2 \sin^{2} \Theta
	\end{equation}
	The above construction is based on an undefined function $f(r,\theta)$ from the equation (\ref{def_R}). It ensures that the transformation of coordinates (\ref{Coord_transform}) is a diffeomorphism.
	The first few terms of the solution (\ref{def_R}) have the following form
	\begin{eqnarray}
		R&=&r+\frac{a^{2} \sin^{2} \theta}{2 r}-\frac{a^{2} m (\cos^{2} \theta -\sin^{2} \theta)}{r^{2}}-\frac{a^{4} \sin^{4} \theta}{8 r^{3}} \nonumber \\
		& &\Scale[0.95]{+\frac{a^{4} m (2 \cos^{4} \theta+\cos^{2} \theta -1)}{2 r^4}+\frac{a^{4} m^2 (4 \cos^{4} \theta+4\cos^{2} \theta -1)-\frac{a^{6}}{8}(\cos^{6} \theta-3 \cos^{4} \theta+3\cos^{2} \theta -1)}{2 r^5}} \label{R_series} \\
		& &\Scale[0.75]{+\frac{a^{4} m[(-14)\cos^{6} \theta+(11 a^{2}+96 m^{2}) \cos^{4} \theta-8 a^{2} \cos^{2} \theta +3 a^{2}]}{8 r^6}+O \left(\frac{1}{r^7} \right)}
		\nonumber
	\end{eqnarray}
	We restrict ourselves to the first seven terms of the series. The lower-order terms have higher precision than is needed to examine the Hamiltonian in our case. Next, we perform an asymptotic analysis of the Hamiltonian. The accuracy of the Round Sphere condition is
	\begin{equation}
		{}^{(2)} \!R = \frac{2}{R^2}+O \left(\frac{1}{R^9} \right)
	\end{equation}
	The equation (\ref{R_series}) with the equations (\ref{sin_Th}) together form the map
	\begin{eqnarray}
		F:(r_{+},\infty)\times(0,\pi) &\to& (R_{+},\infty)\times(0,\pi) \nonumber \\
		F(r,\theta) &\mapsto&(R, \Theta)
	\end{eqnarray}
	where $r_{+},R_{+}$ are, respectively, the outer horizon radius in $r$ and $R$ coordinates. The above map is a homomorphism\footnote{Technically, the map $F$ is restricted by the precision (number of terms) of the equation (\ref{R_series}). The inverse of $F$ can be calculated with the same precision as $F$.}. The inverse of $F$ is the following
	\begin{eqnarray}
		r &=& R-\frac{a^2 \sin^2(\Theta)}{2 R}+\frac{a^2 m(\cos^2 \Theta-\sin^2 \Theta)}{R^2}-\frac{a^4(5 \cos^4 \Theta-6 \cos^2 \Theta+1)}{8 R^3}  \\
		& & \Scale[0.95]{-\frac{ a^4 m( 7\cos^4 \Theta-7 \cos^2 \Theta+1)}{R^4}+\frac{\frac{a^{6}}{16}(21 \cos^6 \Theta-35 \cos^4 \Theta+15 \cos^2 \Theta-1)-a^{4} m^{2} (14 \cos^{4} \Theta-10 \cos^{2} \Theta)}{R^5}} \nonumber \\
		& & \Scale[0.75]{+\frac{a^{4} m [26 a^{2} \cos^6 \Theta-(39 a^{2}+12 m^{2}) \cos^4 \Theta+15 a^{2} \cos^2 \Theta-a^{2}]}{R^6}+O \left(\frac{1}{R^ 7}\right)} \nonumber\\
		\sin \theta&=&\sin \Theta -\frac{a^2 \sin \Theta \cos^2 \Theta}{2 R^2}-\frac{m a^2 \sin \Theta \cos^2 \Theta}{R^3}+\frac{a^4\sin \Theta (7 \cos^4 \Theta-4 \cos^2 \Theta)}{8 R^4} \hspace{1 cm}  \\
		& &\Scale[0.95]{+\frac{a^4 m\sin \Theta (10 \cos^4 \Theta-5 \cos^2 \Theta)}{2 R^5}-\frac{a^4 \cos^{2} \Theta \sin \Theta [33 a^{2}\cos^4 \Theta-(36 a^2+152 m^{2}) \cos^2 \Theta+ 8 a^{2}+48 m^{2}]}{16 R^6}} \nonumber \\
		& &\Scale[0.75]{-\frac{a^4 m \cos^{2} \Theta \sin \Theta [22 a^{2}\cos^4 \Theta-(22 a^2+12 m^{2}) \cos^2 \Theta+ \frac{35}{8} a^{2}]}{R^7}+O \left(\frac{1}{R^ 8}\right)} \nonumber \\
		\cos \theta &=& \cos \Theta +\Scale[1.25]{\frac{a^2 \sin^2 \Theta \cos \Theta}{2 R^2}+\frac{m a^2 \sin^2 \Theta \cos \Theta}{R^3}+\frac{a^4 \cos \Theta (7 \cos^4 \Theta-10 \cos^{2} \Theta+3)}{8 R^4}} \\
		& &\Scale[0.85]{+\frac{a^4 m \cos \Theta (5 \cos^4 \Theta-7 \cos^{2} \Theta+2)}{ R^5}+\frac{a^4 \cos \Theta [(-33)a^{2}\cos^6 \Theta+(63 a^{2}+152m^{2}) \cos^4 \Theta-(35 a^{2}+192 m^{2}) \cos^{2} \Theta+5 a^{2}+40 m^{2}]}{16 R^6}} \nonumber \\
		& &\Scale[0.7]{-\frac{a^{4} m \cos \Theta [ 22 a^{2} \cos^{6} \Theta - (41 a^{2}+12 m^{2}) \cos^{4} \Theta +(22a^{2}+12 m^{2}) \cos^{2} \Theta+3 a^{2}]}{ R^7}+O \left(\frac{1}{R^ 8}\right)} \nonumber
	\end{eqnarray}	  	 	
	$\sin \Theta$ and $\cos \Theta$ respect trigonometric properties.
	
	The two-dimensional objects which are used to obtain the Hamiltonian dynamics are
	\begin{eqnarray}
		\lambda &=& R^2 \sin \Theta+ O \left(\frac{1}{R^5} \right) \label{s:lambda}\\
		k&=&-\frac{2}{R}+\frac{2 m}{R^2}+\frac{m^2}{R^3}+\frac{m^3+6 m a^2(1-3 \cos^2 \Theta)}{R^4}\\
		& &\Scale[0.95]{ +\frac{5 m^4+24 m^2 a^2 (1-4 \cos^2 \Theta)}{4 R^5}+\frac{m [525 a^{4} \cos^{4} \Theta -(450 a^{4}+132 a^{2} m^{2})\cos^{2} \Theta+45 a^{4}+36 a^{2} m^{2}+7 m^{4}]}{4 R^{6}}} \nonumber \\
		& & \Scale[0.85]{+\frac{m^{2}[486 a^{4} \cos^{4} \Theta-(387 a^{4}+54 a^{2} m^{2}) \cos^{2} \Theta +33 a^{4}+15 a^{2} m^{2}+\frac{21}{8} m^{4}]}{ R^{7}}} \nonumber \\
		& &\Scale[0.7]{-\frac{m [ 6468 a^{6} \cos^{6} \Theta-(8820 a^{6}+7377 a^{4} m^{2}) \cos^{4} \Theta+(2940 a^{6}+5298 a^{4} m^{2}+750 a^{2} m^{4}) \cos^{2} \Theta-140 a^{6}-321 a^{4} m^{2}-210 a^{2} m^{4}-33 m^{6}]}{8 R^{8}}+ O \left(\frac{1}{R^9} \right)} \nonumber \\
		k^{2}&=&\frac{4}{R^2}-\frac{8 m}{R^3}+\frac{24 m a^2 (3 \cos^2 \Theta-1)}{R^5}+\frac{24 m^2 a^2 \cos^2 \Theta}{R^6}\\
		& &\Scale[0.95]{-\frac{15 a^{4} m[35\cos^{4} \Theta-30 \cos^{2} \Theta+3 ]}{R^{7}}-\frac{ a^{4} m^{2}[1095\cos^{4} \Theta-882 \cos^{2} \Theta+51 ]}{R^{8}}+ O \left(\frac{1}{R^9} \right)} \nonumber
	\end{eqnarray}
	\begin{eqnarray}
		\mathbf{Q}&=&-2R \sin \Theta+2 m \sin \Theta \Scale[1.3]{+\frac{m^2 \sin \Theta}{R}+\frac{\sin \Theta [m^3+6 m a^2(1-3 \cos^2 \Theta)]}{R^2}} \hspace{0.9 cm}\\
		& &\Scale[0.85]{ +\frac{\sin \Theta [5 m^4+24 m^2 a^2 (1-4 \cos^2 \Theta)]}{4 R^3}+\frac{m \sin \Theta [525 a^{4} \cos^{4} \Theta -(450 a^{4}+132 a^{2} m^{2})\cos^{2} \Theta+45 a^{4}+36 a^{2} m^{2}+7 m^{4}]}{4 R^{4}}} \nonumber \\
		& & \Scale[0.85]{+\frac{m^{2} \sin \Theta[486 a^{4} \cos^{4} \Theta-(387 a^{4}+54 a^{2} m^{2}) \cos^{2} \Theta +33 a^{4}+15 a^{2} m^{2}+\frac{21}{8} m^{4}]}{ R^{5}}} \nonumber \\
		& &\Scale[0.7]{-\frac{m \sin \Theta [ 6468 a^{6} \cos^{6} \Theta-(8820 a^{6}+7377 a^{4} m^{2}) \cos^{4} \Theta+(2940 a^{6}+5298 a^{4} m^{2}+750 a^{2} m^{4}) \cos^{2} \Theta-140 a^{6}-321 a^{4} m^{2}-210 a^{2} m^{4}-33 m^{6}]}{8 R^{6}}}\Scale[0.95]{+ O \left(\frac{1}{R^7} \right)} \nonumber \\
		\mathbf{Q}_{\varphi}&=&-3 m a \sin^3 \Theta+\Scale[1.2]{\frac{5 m a^3 \sin^3 \Theta (5 \cos^2 \Theta-1)}{2 R^2}+\frac{3 m^2 a^3 \sin^3 \Theta (5 \cos^2 \Theta-1)}{R^3}} \\
		& &\Scale[0.95]{-\frac{21 m a^5 \sin^3 \Theta (21 \cos^4 \Theta- 14 \cos^2 \Theta+1)}{8 R^4}-\frac{9 m^{2} a^5 \sin^3 \Theta (21 \cos^4 \Theta- 14 \cos^2 \Theta+1)}{R^5}} \nonumber \\
		& &\Scale[0.75]{+\frac{3 m a^5 \sin^3 \Theta [1287 a^{2} \cos^6 \Theta-(1485 a^{2}+1176 m^{2}) \cos^4 \Theta+(405 a^{2}+704 m^{2}) \cos^2 \Theta-15 a^{2}-40 m^{2}]}{16 R^6}} \Scale[0.95]{+ O \left(\frac{1}{R^7} \right)} \nonumber \\
		g_{\Theta \Theta}&=&R^2 +O \left(\frac{1}{R^ 5}\right) \label{test_gTT} \\
		g_{\varphi \varphi}&=&R^2 \sin^{2} \Theta +O \left(\frac{1}{R^ 5}\right)
	\end{eqnarray}
	The components of the above tensorial objects that are not mentioned are equal to zero.	  	 	
	The equation (\ref{test_gTT}) is an expected result and a significant test of our calculus. In the two-dimensional case, the Gauss--Codazzi equation fulfills the following conditions:
	\begin{enumerate}
		\item The definition of a Round Sphere requires a constant curvature scalar.
		\item The coordinate system $\{R, \Theta, \varphi\}$ is constructed so that $g_{\varphi \varphi}=R^{2} \sin^{2} \Theta$.
	\end{enumerate}
	It means that the $g_{\Theta \Theta}$ component of the metric is equal to $R^{2}$, which concurs with the result (\ref{test_gTT}).
	In the Kerr case, when $g_{t R}=0$, then the ADM counterpart is related to the ADM momentum by the following relation
	\begin{equation}
		g^{t R}=0 \implies \mathbf{Q}_{A}=- \lambda P^{R}_{A}
	\end{equation}
	The other elements needed to construct our Hamiltonian are
	\begin{eqnarray}
		\nu&=& 1\Scale[1.2]{-\frac{m}{R}-\frac{m^2}{2 R^2} -\frac{m^3-m a^2(3 \cos^2 \Theta-1)}{2 R^3}-\frac{5 m^4+4 m^2 a^2(3 \cos^2 \Theta-1)}{8 R^4}} \\
		& &\Scale[0.95]{-\frac{35 a^{4} m \cos^{4} \Theta -30 a^{2} m(a^{2}+\frac{3}{5}m^{2}) \cos^2 \Theta-3 a^{4} m +2 a^{2} m^{3}- 7 m^{5}}{8 R^5}} \nonumber \\
		& &\Scale[0.95]{-\frac{50 a^{4} m^{2} \cos^{4} \Theta - 15 a^{2} m^{2}(2 a^{2}+m^{2})\cos^2 \Theta+2 a^{4} m^{2}-a^{2} m^{4}+\frac{21}{4}m^{6}}{4 R^6}} \Scale[0.95]{+ O \left(\frac{1}{R^7} \right)} \nonumber
	\end{eqnarray}
	\begin{eqnarray}
		\frac{\nu}{k}&=& -\frac{R}{2}+\frac{5 m a^2 \left(3 \cos^2 \Theta-1 \right)}{4 R^2}+\frac{3 m^2 a^2 \left(3 \cos^2 \Theta-1 \right)}{4 R^3}\\
		& &\Scale[0.95]{-\frac{ 245 a^{4} m \cos^{4} \Theta -a^{2} m(210 a^{2}+144 m^{2}) \cos^2 \Theta+21 a^{4} m +48 a^{2} m^{3}}{8 R^4}}\nonumber \\
		& &\Scale[0.75]{-\frac{ 1437 a^{4} m^{2} \cos^{4} \Theta -a^{2} m^{2}(1134 a^{2}+288 m^{2}) \cos^2 \Theta+115 a^{4} m^{2} +96 a^{2} m^{4}}{8 R^5}} \Scale[0.95]{+ O \left(\frac{1}{R^6} \right)} \nonumber \\
		\nu^{\varphi}&=&\Scale[1.2]{-\frac{2 ma}{R^3}+\frac{ma^3(5 \cos^2 \Theta-1)}{R^5}+\frac{2 m^2 a^3(4 \cos^2 \Theta-1)}{R^6}} \\
		& &\Scale[0.95]{-\frac{3 a^5 m (21 \cos^4 \Theta -14 \cos^2 \Theta+1)}{4 R^7}-\frac{2 a^5 m^2 (33 \cos^4 \Theta -22 \cos^2 \Theta+2)}{R^8}+ O \left(\frac{1}{R^9} \right)} \nonumber
	\end{eqnarray}
	The multipole expansion of $\frac{\nu}{k}$ is equal to
	\begin{eqnarray}
		\left(\frac{\nu}{k} \right)_{m}&=&-\frac{R}{2}-\frac{61 a^{4} m^{2}}{20 R^{5}}+ O \left(\frac{1}{R^6} \right) \label{s:nu_k_m}\\
		\left(\frac{\nu}{k} \right)_{d}&=&0+ O \left(\frac{1}{R^6} \right) \\
		\left(\frac{\nu}{k} \right)_{w}&=&(3 \cos^2 \Theta-1) { \left[\frac{5 a^{2} m}{4R^{2}}+\frac{3 a^{2} m^{2}}{4R^{3}}+\frac{6 a^{2} m^{3}}{R^{4}} \right.} \\
		& &{\left.+\frac{1}{R^{5}}\left(12 a^{2} m^{4}-\frac{57 a^{4} m^{2}}{14}\right)\right]} \nonumber \\
		& &\Scale[0.9]{+ (35\cos^{4} \Theta-30 \cos^{2} \Theta+3)}\left[-\frac{7 a^{4} m}{8 R^{4}}-\frac{1437 a^{4} m^{2}}{35 R^5}\right]+ O \left(\frac{1}{R^6} \right) \nonumber
	\end{eqnarray}
	The square of mean curvature splits into multipoles
	\begin{eqnarray}
		(k^{2})_{m}&=&\frac{4}{R^2}-\frac{8 m}{R^3}+\frac{8 m^{2} a^{2}}{R^{6}}+\frac{24 a^{4} m^{2}}{R^{8}}+ O \left(\frac{1}{R^9} \right) \label{s:k2_m} \\
		(k^{2})_{d}&=&0+ O \left(\frac{1}{R^9} \right) \\
		(k^{2})_{w}&=&(3 \cos^2 \Theta-1)\left[\frac{24 m a^{2}}{R^{5}}+\frac{8 a^{2} m^{2}}{R^{6}}-\frac{132 a^{4} m^{2}}{7 R^{8}}\right] \nonumber \\
		& &+\Scale[0.9]{(35\cos^{4} \Theta-30 \cos^{2} \Theta+3)} \left[-\frac{15 a^{4} m}{R^{7}}-\frac{219 a^{4} m^{2}}{7 R^{8}}\right]+ O \left(\frac{1}{R^9} \right)
	\end{eqnarray}
	The decompositions of $\nu^{A}$ and $\mathbf{Q}_{A}$ into its longitudinal and transversal parts are
	\begin{eqnarray}
		\left({\! \hphantom{|}}_{|| \!} \nu \right)^{\varphi} &=&0 \\
		\left({\! \hphantom{|}}_{\perp \!} \nu \right)^{\varphi} &=&\nu^{\varphi} \\
		({\! \hphantom{|}}_{|| \!} \mathbf{Q})_{\varphi}&=&0 \\
		( {\! \hphantom{|}}_{\perp \! \!} \mathbf{Q})_{\varphi}&=&\mathbf{Q}_{\varphi}
	\end{eqnarray}
	The multipole expansions of $\left({\! \hphantom{|}}_{\perp \!} \nu \right)^{\varphi}$ and $( {\! \hphantom{|}}_{\perp \! \!} \mathbf{Q})_{\varphi}$
	\begin{eqnarray}
		{\! \hphantom{|}}_{\perp \!} \nu_{ d}^{\varphi} &=&-\frac{2 m a}{R^{3}}-\frac{2 m^{2} a^{3}}{5 R^{6}}-\frac{6 a^{5} m^{2}}{7 R^{8}} + O \left(\frac{1}{R^9} \right) \label{s:nu_phi_d} \\
		{\! \hphantom{|}}_{\perp \!} \nu_{ w}^{\varphi} &=& \Scale[0.95]{\left(5 \cos^2 \Theta-1\right)} \left[\frac{m a^{3}}{R^{5}}+\frac{8 m^{2} a^{3}}{5 R^{6}} \right] \\
		& & \Scale[0.95]{+\left(21 \cos^4 \Theta-14 \cos^2 \Theta+1\right)} \Scale[1.2]{\left[-\frac{3 m a^{5}}{4 R^{7}}-\frac{-22 m^{2} a^{5}}{7 R^{8}}\right]+ O \left(\frac{1}{R^9} \right)} \nonumber \\
		( {\! \hphantom{|}}_{\perp \! \!} \mathbf{Q}_{\varphi})_{d}&=&-\sin^{3} \Theta \left(3 m a\right) + O \left(\frac{1}{R^7} \right) \label{s:Qbf_d}\\
		( {\! \hphantom{|}}_{\perp \! \!} \mathbf{Q}_{\varphi})_{w}&=&\Scale[0.85]{\sin^{3} \Theta \left(5 \cos^2 \Theta-1\right)} \left[\frac{5 m a^{3}}{2 R^{2}}+\frac{3 m^{2} a^{3}}{R^{3}}-\frac{3 m^3 a^{5}}{R^{6}}\right] \\
		& &\Scale[0.85]{+\sin^{3} \Theta \left(21 \cos^4 \Theta-14 \cos^2 \Theta+1\right)} \left[-\frac{21 m a^{5}}{8 R^{4}}-\frac{9 m^{2} a^{5}}{R^{5}}-\frac{21m^{3} a^{5}}{2R^{6}} \right] \nonumber \\
		& &+\Scale[0.85]{\sin^{3} \Theta \left( 429 \cos^{6} \theta-495 \cos^{4} \theta+135 \cos^{2} \theta-5 \right)} \frac{9 m a^{7}}{16 R^{6}} \nonumber + O \left(\frac{1}{R^7} \right)
	\end{eqnarray}
	Using the series (\ref{s:lambda}), (\ref{s:nu_k_m}), (\ref{s:k2_m}), (\ref{s:nu_phi_d}), (\ref{s:Qbf_d}), the integrals of the two parts: the scalar part and the vector part of the Hamiltonian are
	\begin{eqnarray}
		\int_{0}^{\pi} \mathrm{d} \Theta \lambda  \left(\frac{\nu}{k} \right)_{m} (k^2_{m})&=&-4 R+8 m -\frac{8 m^{2} a^{2}}{R^{3}}-\frac{242 m^{2} a^{4}}{5 R^{5}} + O \left(\frac{1}{R^6} \right) \\
		\int_{0}^{\pi} \mathrm{d} \Theta \left({\! \hphantom{|}}_{\perp \!} \nu_{ A} \right)_{d} {\! \hphantom{|}}_{\perp \! \!} \mathbf{Q}^{A}_{d}&=&\frac{8 m^{2} a^{2}}{R^{3}}+ O \left(\frac{1}{R^6} \right)
	\end{eqnarray}
	The mean curvature of the reference frame is very simple: We embed a surface with an almost spherical metric into flat spacetime
	\begin{equation}
		k_{0}=-\frac{2}{R}+ O \left(\frac{1}{R^9} \right)
	\end{equation}
	It is a pure monopole function. The above enables one to compute an asymptotic series of the Hamiltonian (\ref{Hamiltonian_multipoled_final}) for the Round Sphere of radius $R$
	\begin{equation}
		\mathbf{H}_{\partial \mathcal{V}}= m-\frac{242 m^{2} a^{4}}{40 R^{5}} + O \left(\frac{1}{R^6} \right) \label{s:Hamiltonian_round}
	\end{equation}
	The asymptotic Hamiltonian for the Round Sphere (\ref{s:Hamiltonian_round}) has the following properties:
	\begin{enumerate}
		\item The mass term $m$ for Kerr spacetime is exposed in comparison with the first unwanted deviation term for big values of radial coordinate\footnote{For example: using $m=a$ and $R=4 m$, we obtain the value of the first deviation term equal to $-0.006 m$.} $R$.
		\item The standard deviation related to the angular momentum and its square does not occur. It results from the three facts: specific choice of selected surface -- Round Sphere, the restriction only to the lower-rank multipoles in the Hamiltonian, and the correction which comes from the $\nu^{A}\mathbf{Q}_{A}$ term.
		\item The deviation leading term has a minus sign. The mass is an increasing function of the radial coordinate $R$.
	\end{enumerate}
	\subsection{Results for Rigid Spheres}
	\label{ssec:Results_Rigid}
	
	The definition of Rigid Spheres leads to an elliptic equation. The solution is a two-argument function which is constant on the surface $\mathcal{S}$
	\begin{equation}
		R=p(r,\theta) \label{def_R_Rigid}
	\end{equation}
	The first few terms of $R$ are in the form of an asymptotic series:
	\begin{eqnarray}
		R&=&r+\Scale[1.05]{\frac{a^{2} \sin^{2} \theta}{2 r}-\frac{a^{2} m (5\cos^{2} \theta - 1 )}{2 r^{2}} - \frac{a^{2}[a^{2} \cos^{4} \theta+(42 m^2-2 a^2) \cos^{2} \theta+a^2-6 m^{2}]}{8 r^{3}}} \label{R_rigid_series} \\
		& &\Scale[0.75]{+\frac{a^{2} m [65 a^2\cos^{4} \theta +(189 m^2-90 a^2) \cos^{2} \theta+9 a^2-135 m^2]}{24 r^4}-\frac{a^2 [ 9 a^4 \cos^6 \theta + (2794 a^2 m^2-27 a^4) \cos^4 \theta + (27 a^4 - 1896 m^2 a^2 + 1701 m^4  )\cos^{2} \theta-9 a^4 + 222 m^2 a^2+729 m^4]}{144 r^5}} \nonumber \\
		& &\Scale[0.75]{+\frac{m a^2 [ 12078 a^4 \cos^6 \theta + (339340 a^2 m^2 - 29250 a^4) \cos^4 \theta + (20250 a^4 - 177000 m^2 a^2 + 76545 m^4) \cos^2 \theta -1350 a^4+ 37140 m^2 a^2 - 142155 m^4 ]}{4320 r^6}} \Scale[1]{+O \left(\frac{1}{r^7} \right)} \nonumber
	\end{eqnarray}
	We restrict ourselves to the first seven terms of the above series. The lower-order terms have higher precision than is needed to examine the Hamiltonian in our case. We perform an asymptotic analysis of the Hamiltonian. The accuracy of the Rigid Sphere condition is
	\begin{eqnarray}
		k&=&\Scale[1.1]{-\frac{2}{R}+\frac{2 m}{R^2}+\frac{m^2}{R^3}+\frac{m^3}{R^4} +\frac{5 m^4}{4 R^5}+\frac{7 m^{5}}{4 R^{6}}+\frac{21 m^6}{8 R^7}+\frac{33 m^7}{8 R^8} + O \left(\frac{1}{R^9} \right)} \\
		k^2&=&\frac{4}{R^2}-\frac{ 8 m}{R^3} + O \left(\frac{1}{R^{10}} \right)
	\end{eqnarray}
	
	Rigid Spheres do not possess a natural multipole structure. We receive it by introducing the Conformally Spherical coordinates (see (\ref{conformal_transform})). The coordinates are described in detail in \cite{Leski2013_Rigid_Spheres}.
	\paragraph{Conformally spherical coordinates}
	Let $h_{AB}$ be an induced metric on $\mathcal{S}$. It has the form:
	\begin{equation}
		h_{AB} \mathrm{d} x^{A} \mathrm{d}x^{B}=h_{\theta \theta} \mathrm{d} \theta^2+h_{\varphi \varphi} \mathrm{d} \varphi^2
	\end{equation}
	We transform coordinates into $(\Theta, \varphi)$, so the transformed metric on $\mathcal{S}$ is conformally spherical, and the transformation preserves axial symmetry for the $\varphi$ coordinate. It leads to the relation
	\begin{equation}
		h_{\theta \theta} \mathrm{d} \theta^2+h_{\varphi \varphi} \mathrm{d} \varphi^2=\Psi^{2} (\Theta) \left[\mathrm{d} \Theta^2+ \sin^2 \Theta \mathrm{d} \varphi^2 \right] \label{conformal_transform}
	\end{equation}
	which holds if and only if
	\begin{eqnarray}
		\sin \Theta&=&\frac{1}{\cosh A}=\frac{2 }{\mathrm{e}^{A}+\mathrm{e}^{-A}} \label{sin_Th_eqlib}
	\end{eqnarray}
	where
	\begin{equation}
		A=\int \sqrt{\frac{h_{\theta \theta}(R,\theta)}{h_{\varphi \varphi}(R,\theta)}} \mathrm{d} \theta
	\end{equation}
	The transformation is valid for all $A \in \mathbb{R}_{+}$.
	It is convenient to perform a three-dimensional transformation of coordinates
	\begin{equation}
		(r,\theta,\varphi) \to (R, \Theta,\varphi) \label{Coord_transform_equilibr}
	\end{equation}
	where $R \in \mathbb{R}_{+}$ is implicitly given by the equation (\ref{def_R_Rigid}).  $\Theta \in (0,\pi)$ is defined by equations (\ref{sin_Th_eqlib}).
	They fulfill basic trigonometric properties, i.e., $\sin^{2} \Theta+\cos^{2} \Theta=1$. $\varphi$ is transformed identically. It turns out that the coordinates $(\Theta,\varphi)$ from transformation (\ref{Coord_transform_equilibr}) are Conformally Spherical and also Equlibrated coordinates (see definition \ref{df_Equilibrated} from the appendix \ref{multipole_expansion} ). They possess the necessary structure to introduce a multipole decomposition on Rigid Spheres.
	The transformation (\ref{Coord_transform_equilibr}) can be treated like a map
	\begin{eqnarray}
		F:(r_{+},\infty)\times(0,\pi)\times(0,2 \pi) &\to& (R_{+},\infty)\times(0,\pi) \times(0,2 \pi) \nonumber \\
		F(r,\theta,\varphi) &\mapsto&(R, \Theta,\varphi)
	\end{eqnarray}
	where $r_{+},R_{+}$ are respectively outer horizon radiuses in $r$ and $R$ coordinates. The above map is a homomorphism\footnote{Technically, the map $F$ is restricted by the precision (number of terms) of the equation (\ref{R_series}). The inverse of $F$ can be calculated with the same precision as $F$.}. The inverse of $F$ is the following
	\begin{eqnarray}
		r &=& R-\Scale[1.1]{\frac{a^2 \sin^2(\Theta)}{2 R}-\frac{a^2 m(5 \cos^2 \Theta-1)}{R^2}-\frac{a^2[5 a^2 \cos^4 \Theta-6(a^2+7 m^2) \cos^2 \Theta+a^2+6 m^2]}{8 R^3}}  \\
		& & \Scale[0.95]{+\frac{ a^2 m( 61 a^2\cos^4 \Theta-(54 a^2+189 m^2) \cos^2 \Theta+9 a^2+135 m^2)}{24 R^4}} \nonumber \\
		& & \Scale[0.75]{+\frac{a^{2}[21 a^4 \cos^6 \Theta-35 a^2 \left(a^2+\frac{554}{315} m^2\right) \cos^4 \Theta+\frac{135 a^4+588 a^2 m^2+1701 m^4}{9} \cos^2 \Theta-a^4-\frac{22}{3} a^2 m^2+81 m^4]}{16 R^5}} \label{eq:rRrigid}\\
		& &\Scale[0.75]{-\frac{m a^2 [27198 a^4 \cos^6 \Theta -(35730 a^4 +17870 a^2 m^2) \cos^4 \Theta +(11610 a^4 +54660 m^2 a^2+76545 m^4) \cos^2 \Theta-1350 a^4-31710 m^2 a^2-142155 m^4]}{4320 R^6}}\Scale[1]{+O \left(\frac{1}{R^ 7}\right)}\nonumber \\
		\sin \theta&=&\sin \Theta -\frac{a^2 \sin \Theta \cos^2 \Theta}{2 R^2}-\frac{m a^2 \sin \Theta \cos^2 \Theta}{R^3}+\frac{a^4\sin \Theta (7 \cos^4 \Theta-4 \cos^2 \Theta)}{8 R^4} \hspace{1 cm}  \\
		& &\Scale[0.95]{+\frac{a^4 m\sin \Theta (\cos^4 \Theta-2 \cos^2 \Theta)}{2 R^5}-\frac{a^4 \cos^{2} \Theta \sin \Theta [33 a^{2}\cos^4 \Theta-(36 a^2+92 m^{2}) \cos^2 \Theta+ 8 a^{2}-12 m^{2}]}{16 R^6}} \nonumber \\
		& &\Scale[0.75]{+\frac{a^4 m \cos^{2} \Theta \sin \Theta [25 a^{2}\cos^4 \Theta+(36 a^2-243 m^{2}) \cos^2 \Theta-18 a^{2}+81 m^2]}{24 R^7}}+\Scale[1]{O \left(\frac{1}{R^ 8}\right)} \nonumber \\
		\cos \theta &=& \cos \Theta +\Scale[1.25]{\frac{a^2 \sin^2 \Theta \cos \Theta}{2 R^2}+\frac{m a^2 \sin^2 \Theta \cos \Theta}{R^3}+\frac{a^4 \cos \Theta (7 \cos^4 \Theta-10 \cos^{2} \Theta+3)}{8 R^4}} \\
		& &\Scale[0.85]{+\frac{a^4 m \cos \Theta ( \cos^4 \Theta-2 \cos^{2} \Theta+1)}{ 2 R^5}-\frac{a^4 \cos \Theta [33 a^{2}\cos^6 \Theta-(63 a^{2}+92m^{2}) \cos^4 \Theta+(35 a^{2}+72 m^{2}) \cos^{2} \Theta-5 a^{2}+20 m^{2}]}{16 R^6}} \label{eq:costhrigid} \\ 				  	 
		& &\Scale[0.7]{+\frac{a^{4} m \cos \Theta [ 25 a^{2} \cos^{6} \Theta - (7 a^{2}+243 m^{2}) \cos^{4} \Theta -(21a^{2}-324 m^{2}) \cos^{2} \Theta+3 a^{2}-81 m^2]}{ R^7}} \Scale[1]{+O \left(\frac{1}{R^ 8}\right)} \nonumber\\
		\varphi&=&\varphi
	\end{eqnarray}	 
	The two-dimensional objects which are used to obtain the Hamiltonian dynamics are
	\begin{eqnarray}
		g_{\Theta \Theta}&=&\Psi^2 \label{gTT_rigid} \\
		g_{\varphi \varphi}&=&\Psi^2 \sin^{2} \Theta \\
		\lambda &=& \Psi^2 \sin \Theta \label{s:lambda_rigid}
	\end{eqnarray}
	where
	\begin{eqnarray}
		\Psi^2&=&R^2-\frac{3 a^2 m(3 \cos^2 \Theta-1)}{R}+\frac{3a^2 m^2 (7 \cos^2 \Theta-1)}{2 R^2}  \label{conformal_factor} \\
		& &\Scale[0.85]{+\frac{m a^2 [175 a^2 \cos^{4} \Theta - (150 a^2+189 m^2) \cos^2 \Theta+15 a^2+135 m^2]}{12 R^3}+\frac{m^2 a^2 [1354 a^2 \cos^{4} \Theta - (96 a^2-1701 m^2) \cos^2 \Theta - 66 a^2+729 m^2]}{72 R^4}} \nonumber \\
		& &\Scale[0.75]{+\frac{1}{R^5} \left[-\frac{1617}{40} m a^6 \cos^6 \Theta+\left(\frac{441}{8}m a^6 -\frac{1103}{27} m^3 a^4 \right) \cos^{4} \Theta  -\left(\frac{147}{8} m a^6 +\frac{587}{36} m^3 a^4 -\frac{567}{16} a^2 m^5\right)\cos^2 \Theta \right.} \nonumber \\
		& & \Scale[0.75]{\left.+\frac{7}{8} m a^6+\frac{353}{36} m^3 a^4+\frac{1053}{16} m^5 a^2 \right]} \Scale[1]{+O \left(\frac{1}{R^ 6}\right)} \nonumber
	\end{eqnarray}
	\begin{eqnarray}
		\mathbf{Q}_{\varphi}&=&-3 m a \sin^3 \Theta+\Scale[1.0]{\frac{5 m a^3 \sin^3 \Theta (5 \cos^2 \Theta-1)}{2 R^2}+\frac{3 m^2 a^3 \sin^3 \Theta (5 \cos^2 \Theta-1)}{R^3}} \Scale[0.95]{+ O \left(\frac{1}{R^4} \right)}
	\end{eqnarray}
	The other elements needed to construct our Hamiltonian are
	
	\begin{eqnarray}
		{}^{(2)}R&=&\frac{2}{R^2}-\frac{12 a^2 m (3 \cos^2 \Theta -1)}{R^5}+\frac{6 a^2 m^2 (7 \cos^2 \Theta-3)}{R^6}+ O \left(\frac{1}{R^7} \right) \\
		\nu&=& 1\Scale[1.2]{-\frac{m}{R}-\frac{m^2}{2 R^2} -\frac{m^3-m a^2(3 \cos^2 \Theta-1)}{2 R^3}-\frac{5 m^4+8 m^2 a^2(3 \cos^2 \Theta-2)}{8 R^4}} \\
		& &\Scale[0.95]{-\frac{35 a^{4} m \cos^{4} \Theta -30 a^{2} m(a^{2}+\frac{4}{5}m^{2}) \cos^2 \Theta-3 a^{4} m +8a^{2} m^{3}- 7 m^{5}}{8 R^5}} \Scale[0.95]{+ O \left(\frac{1}{R^6} \right)} \nonumber\\
		\frac{\nu}{k}&=& -\frac{R}{2}+\frac{ m a^2 \left(1-3 \cos^2 \Theta \right)}{4 R^2}+\frac{3 m^2 a^2 \left(\cos^2 \Theta-1 \right)}{4 R^3}\\
		& &\Scale[0.95]{+\frac{m a^2 \left[ 35 a^2 \cos^{4} \Theta -(30 a^2 +18 m^2) \cos^2 \Theta+3 a^2 -18 m^2\right]}{16 R^4}}\nonumber \\
		& &\Scale[0.95]{-\frac{m^2 a^2 \left[ 256 a^2 \cos^{4} \Theta -(204 a^2+81 m^2) \cos^2 \Theta+48 a^2+243 m^2\right]}{48 R^5}} \Scale[0.95]{+ O \left(\frac{1}{R^6} \right)} \nonumber \\
		\nu^{\varphi}&=& \Scale[1.2]{-\frac{2 ma}{R^3}+\frac{ma^3(5 \cos^2 \Theta-1)}{R^5}-\frac{ m^2 a^3(19 \cos^2 \Theta-7)}{R^6}} \\
		& &-\Scale[0.95]{\frac{3 a^3 m \left[21 a^2 \cos^4 \Theta - (14 a^2+42 m^2) \cos^2 \Theta+a^2+6 m^2 \right]}{4 R^7}}\Scale[1]{+ O \left(\frac{1}{R^8} \right)} \nonumber
	\end{eqnarray}		
	The components of the above tensorial objects that are not mentioned are equal to zero.  	 	
	\paragraph{The mean curvature of the reference frame}
	The mean curvature of the reference frame is the mean curvature of $\mathcal{S}$ equipped with an induced metric $h_{AB}$ embedded into a three-dimensional Euclidean space. It is not apparent how to transform this three-dimensional metric, with a two-dimensional conformally spherical part, into the three-dimensional Euclidean metric. We propose the following method:
	let us consider the transformation of coordinates in the form
	\begin{equation}
		x=\Psi(R, \Theta) \sin \Theta \cos \varphi \quad y=\Psi(R, \Theta) \sin \Theta \sin \varphi \quad z=G(R,\Theta) \label{reference_coords_transform}
	\end{equation}
	with the condition
	\begin{equation}
		G(R,\Theta)=\int \sqrt{\Psi^{2} (R, \Theta)-\left[\frac{\partial \Psi(R, \Theta) }{\partial \Theta} \sin \Theta+ \Psi (R, \Theta) \cos \Theta\right]^2} \mathrm{d} \Theta
	\end{equation}
	It leads to the following transformation of the flat metric
	\begin{eqnarray}
		\mathrm{d}x^2 + \mathrm{d}y^2 + \mathrm{d} z^2= E_{R R} \mathrm{d} R^2 + 2 E_{R \Theta}  \mathrm{d} R \mathrm{d} \Theta+ \Psi^{2} (\Theta) \left[\mathrm{d} \Theta^2+ \sin^2 \Theta \mathrm{d} \varphi^2 \right]
	\end{eqnarray}
	The mean curvature of the surface $\mathcal{S}=\{R=\mathrm{const.}\}$ embedded in three-dimensional Euclidean space is
	\begin{equation}
		k_{0}=- \frac{g^{i j} \Gamma^{R}_{ij}(E_{kl})}{\sqrt{E^{RR}}}
	\end{equation}
	In our case, if we state that the conformal factor in the transformation (\ref{reference_coords_transform}) is equal to (\ref{conformal_factor}), we receive the following results
	\begin{eqnarray}
		E_{RR}&=&1+\frac{6 a^2 m (3 \cos^2 \Theta-1)}{R^3}-\frac{9 a^2 m^2(7 \cos^2 \Theta -1)}{2  R^4}\\
		& &-\Scale[0.95]{\frac{a^2 m[175 a^2 \cos^4 \Theta +(150 a^2+189 m^2) \cos^2 \Theta+ 15 a^2 +135 m^2]}{3 R^5} + O \left(\frac{1}{R^6}\right)} \nonumber \\
		E_{R \Theta}&=&\Scale[0.95]{\frac{9 a^2 m \sin \Theta \cos \Theta}{R^2}-\frac{21 a^2 m^2 \sin \Theta \cos \Theta}{2 R^3}-\frac{175 a^2 m \sin \Theta \cos \Theta \left[a^2 \cos^2 \Theta-\frac{3}{7} a^2-\frac{27 }{50} m^2\right]}{6 R^4}} \\
		& & \Scale[0.75]{+ \frac{a^2 m^2 \sin \Theta \cos \Theta\left[4096 a^2 \cos^2 \Theta - 2820 a^2-1701 m^2 \right]}{ 72 R^5}}  \Scale[0.95]{+ O \left(\frac{1}{R^6}\right)} \nonumber
	\end{eqnarray}
	\begin{eqnarray}
		k_{0}&=&-\frac{2}{R}+\frac{6 a^2 m (3 \cos^{2}-1)}{R^4}-\frac{3 a^2 m^2 (7 \cos^2 \Theta-3)}{R^5} \\
		& &-\Scale[0.95]{\frac{3 a^2 m \left[175 a^2 \cos^4 \Theta-(150 a^2+42 m^2) \cos^2 \Theta+15 a^2+6 m^2 \right]}{4 R^6}+ O \left(\frac{1}{R^7}\right)} \nonumber
	\end{eqnarray}
	The multipole expansions of selected objects are
	\begin{eqnarray}
		(k^{2})_{m}&=&\frac{4}{R^2}-\frac{ 8 m}{R^3} + O \left(\frac{1}{R^{10}} \right) \label{s:k2_rigid} \\
		(k^{2})_{d}&=&0+ O \left(\frac{1}{R^{10}} \right) \\
		(k^{2})_{w}&=&0+ O \left(\frac{1}{R^{10}} \right)
	\end{eqnarray}
	
	\begin{eqnarray}
		\left(\frac{\nu}{k} \right)_{m}&=&-\frac{R}{2}-\frac{m^2 a^2}{2 R^3}-\frac{3 m^3 a^2}{2 R^{4}}-\left[\frac{13 m^2 a^4}{20}+\frac{9 m^4 a^2}{2}\right]\frac{1}{R^5}+O \left(\frac{1}{R^6} \right) \label{s:nu_k_m_rigid}\\
		\left(\frac{\nu}{k} \right)_{d}&=&0+ O \left(\frac{1}{R^6} \right) \\
		\left(\frac{\nu}{k} \right)_{w}&=&(3 \cos^2 \Theta-1) { \left[-\frac{m a^2}{4 R^2}+\frac{m^2 a^2}{4 R^3}-\frac{3 m^3 a^2}{8 R^4}-\frac{3 m^2 a^4}{28 R^5}+\frac{9 m^4 a^2}{16 R^5} \right]} \\
		& &+\Scale[0.9]{ (35\cos^{4} \Theta-30 \cos^{2} \Theta+3)}\left[\frac{m a^4}{16 R^4}-\frac{16 m^2 a^4}{105 R^5}\right]+ O \left(\frac{1}{R^6} \right) \nonumber
	\end{eqnarray}
	The decompositions of $\nu^{A}$ and $\mathbf{Q}_{A}$ into its longitudinal and transversal part are
	\begin{eqnarray}
		\left({\! \hphantom{|}}_{|| \!} \nu \right)^{\varphi} &=&0 \\
		\left({\! \hphantom{|}}_{\perp \!} \nu \right)^{\varphi} &=&\nu^{\varphi} \\
		({\! \hphantom{|}}_{|| \!} \mathbf{Q})_{\varphi}&=&0 \\
		( {\! \hphantom{|}}_{\perp \! \!} \mathbf{Q})_{\varphi}&=&\mathbf{Q}_{\varphi}
	\end{eqnarray}
	The multipole expansions of $\left({\! \hphantom{|}}_{\perp \!} \nu \right)^{\varphi}$ and $( {\! \hphantom{|}}_{\perp \! \!} \mathbf{Q})_{\varphi}$ are
	\begin{eqnarray}
		{\! \hphantom{|}}_{\perp \!} \nu_{ d}^{\varphi} &=&-\frac{2 m a}{R^{3}}+\frac{16 m^2 a^3}{5 R^6}+\frac{9 m^3 a^3}{5 R^7}+ O \left(\frac{1}{R^8} \right) \label{s:nu_phi_d_rigid} \\
		{\! \hphantom{|}}_{\perp \!} \nu_{ w}^{\varphi} &=& \Scale[0.95]{\left(5 \cos^2 \Theta-1\right)} \left[\frac{m a^{3}}{R^{5}}-\frac{19 m^2 a^3}{5 R^6}+\frac{63 m^3 a^3}{10 R^7} \right] \\
		& & \Scale[0.95]{+\left(21 \cos^4 \Theta-14 \cos^2 \Theta+1\right)} \Scale[1.2]{\left[-\frac{3 m a^5}{4 R^7}\right]+ O \left(\frac{1}{R^8} \right)} \nonumber \\
		( {\! \hphantom{|}}_{\perp \! \!} \mathbf{Q}_{\varphi})_{d}&=&\sin^{3} \Theta \left(-3 m a\right) + O \left(\frac{1}{R^4} \right) \label{s:Qbf_d_rigid}\\
		( {\! \hphantom{|}}_{\perp \! \!} \mathbf{Q}_{\varphi})_{w}&=&\Scale[0.85]{\sin^{3} \Theta \left(5 \cos^2 \Theta-1\right)} \left[\frac{5 m a^{3}}{2 R^{2}}+\frac{3 m^{2} a^{3}}{R^{3}}\right]+ O \left(\frac{1}{R^4} \right)
	\end{eqnarray}
	and the square of the mean curvature of the reference frame is
	\begin{eqnarray}
		(k^{2}_{0})_{m}&=&\frac{4}{R^2}-\frac{8 m^2 a^2}{R^6}-\frac{24 m^{3} a^{2}}{R^{7}}+ O \left(\frac{1}{R^8} \right) \label{s:k2_m_rigid} \\
		(k^{2}_{0})_{d}&=&0+ O \left(\frac{1}{R^9} \right) \\
		(k^{2}_{0})_{w}&=&(3 \cos^2 \Theta-1)\left[-\frac{24 m a^{2}}{R^{5}}+\frac{28 a^{2} m^{2}}{R^{6}}-\frac{42 a^{2} m^{3}}{ R^{7}}\right] \nonumber \\
		& &+\Scale[0.9]{(35\cos^{4} \Theta-30 \cos^{2} \Theta+3)} \left[\frac{15 a^{4} m}{R^{7}}\right]+ O \left(\frac{1}{R^8} \right)
	\end{eqnarray}

	Using the series
	(\ref{s:lambda_rigid}), (\ref{s:nu_k_m_rigid}), (\ref{s:k2_m_rigid}), (\ref{s:nu_phi_d_rigid}) and (\ref{s:Qbf_d_rigid}),
	the integrals of the two parts, the scalar part and the vector part of the Hamiltonian, are
	\begin{eqnarray}
		\int_{0}^{\pi} \mathrm{d} \Theta \lambda  \left(\frac{\nu}{k} \right)_{m} \left[(k^2)_{m}-(k^2_{0})_{m}\right]&=&8 m -\frac{8 m^{2} a^{2}}{R^{3}}+\frac{72 m^{4} a^{2}}{R^{5}} + O \left(\frac{1}{R^6} \right) \\
		\int_{0}^{\pi} \mathrm{d} \Theta \left({\! \hphantom{|}}_{\perp \!} \nu_{ A} \right)_{d} {\! \hphantom{|}}_{\perp \! \!} \mathbf{Q}^{A}_{d}&=&\frac{8 m^{2} a^{2}}{R^{3}}
		+ O \left(\frac{1}{R^6} \right)
	\end{eqnarray}
	The above enables one to compute an asymptotic series of the Hamiltonian (\ref{Hamiltonian_multipoled_final}) for the Rigid Sphere of radius $R$
	\begin{equation}
		\mathbf{H}_{\partial \mathcal{V}}= m +\frac{9 m^{4} a^{2}}{R^{5}} + O \left(\frac{1}{R^6} \right) \label{s:Hamiltonian_rigid}
	\end{equation}
	The asymptotic Hamiltonian for the Round Sphere (\ref{s:Hamiltonian_rigid}) has the following properties:
	\begin{enumerate}
		\item The mass term $m$ for Kerr spacetime is exposed in comparison with the first unwanted deviation term for big values of radial coordinate\footnote{For example: using $m=a$ and $R=4 m$, we obtain the value of the first deviation term equal to $0.009 m$.} $R$.
		\item The standard deviation related to the angular momentum and its square does not occur. It results from the three facts: a specific choice of the selected surface -- a Rigid Sphere, a restriction only to the lower-rank multipoles in the Hamiltonian, and the correction which comes from the $\nu^{A}\mathbf{Q}_{A}$ term.
		\item The deviation leading term has a plus sign. The mass is a decreasing function of the radial coordinate $R$.
	\end{enumerate}
	\subsection{Discussion. Comparison of the obtained results for Round Spheres and Rigid Spheres}
	\label{ssec:Comparsion_Results}

	In the sections \ref{ssec:Results_Round} and \ref{ssec:Results_Rigid} are presented the methods which lead to the energy formulas: (\ref{s:Hamiltonian_round}) for Round Spheres and (\ref{s:Hamiltonian_rigid}) for Rigid Spheres. The above formulas cannot be compared directly because they are not exposed in the same coordinates (Although we use the same symbols --- $R$ and $\Theta$ --- in sections \ref{ssec:Results_Round} and \ref{ssec:Results_Rigid}, they have a different meaning!). The coordinate $\varphi$ and the parameters $m$ and $a$ are the same in the paper.
	
	As the next step, we present how to compare the results and conclude the obtained analysis. Let us note that the condition $\delta g_{AB} = 0$ is not fulfilled in bulk for rigid spheres. However, using equilibrated coordinates (section \ref{multipole_expansion}), the Hamiltonian obtained for rigid spheres in the section \ref{ssec:Results_Rigid} is well defined. Comparison of the Hamiltonians provides essential feedback, which is valuable for the further improvement of the selected surfaces and the presented Hamiltonian dynamics. In particular, better-suited gauge conditions may be set. We highlight a few observations on that matter:
	\begin{enumerate}
		\item For rigid spheres, a coordinate system in which the two-dimensional metric is spherical and simultaneously one of the coordinates covers with the rotational symmetry of Kerr metric does not exist. The conditions for external geometry, which comes from the rigid sphere definition, are incompatible with the requests for inner geometry of the surface, which comes from a specific choice of coordinates. In the other words, if we fix the surface $R=\mathrm{const.}$ by the equation (\ref{R_series}), there do not exist spherical coordinates $(\Theta,\varphi)$ with the metric $f(\Theta) \mathrm{d} \Theta^2+ \sin^{2} \Theta \mathrm{d} \varphi^{2}$ such that $\varphi$ cover with the corresponding Boyer--Lindquist coordinate. It is also important for further theoretical constructions for Rigid Spheres that use the axial symmetry of Kerr spacetime.
		\item The equilibrated coordinates (see definition \ref{df_Equilibrated}) are a good choice of parametrization for investigated surfaces. They ensure the multipole structure and are compatible with the axial symmetry of Kerr spacetime. In the equilibrated coordinates, the metric has one free function --- conformal factor $\Psi$ with the integral condition for its barycenter (see definition \ref{df_Equilibrated}).
		 
		\item The results (\ref{s:Hamiltonian_round}) and (\ref{s:Hamiltonian_rigid}) have the same rank $R^{-5}$ of the unwanted correction to the Hamiltonian. It is expected that a family of topological spheres for which the correction will vanish up to a higher order will exist. We call them mixed geometry spheres. Mixed geometry spheres have a balanced set of conditions for interior and exterior geometry. We do not specify their exact definition now. They will be investigated in a separate paper.
		
        \item  The Hamiltonians derived for round spheres and rigid spheres impose upper and lower bounds on the mass\footnote{The leading terms of corrections to the mass in Hamiltonians have opposite signs in the case of rigid spheres and round spheres.}. This allows one to estimate the mass value in a precise range and the accuracy of the obtained result. Such estimation can be particularly valuable for analyzing experimental data.

	\end{enumerate}
	 
	One of the ways to compare rigid spheres and round spheres is to express rigid sphere radius in terms of Round Sphere radius. The equations \eqref{sin_Th}, \eqref{R_series} and \eqref{eq:rRrigid}-\eqref{eq:costhrigid} can be rearranged into
	\begin{equation}
		r\left(R_{\mathrm{Rigid}},\theta\right)=r\left(R_{\mathrm{Round}},\theta \right) \, ,
	\end{equation}
	which for a fixed $\theta$ form an implicit equation $R_{\mathrm{Rigid}}\left(R_{\mathrm{Round}}\right)$. Setting $\theta=\frac{\pi}{2}$, we find
	\begin{equation}
		\begin{aligned}
			R_{\mathrm{Rigid}}=&R_{\mathrm{Round}}-\frac{3 m a^2}{2 R_{\mathrm{Round}}^2}+\frac{3 m^2 a^2}{4 R_{\mathrm{Round}}^3}-\frac{m a^2 (45 m^2+5 a^2)}{8 R_{\mathrm{Round}}^4}-\frac{m^2 a^2\left(243 m^2+140 a^2\right)}{48 R_{\mathrm{Round}}^5}	\\
			&\Scale[1.1]{-\frac{m a^2\left(9477 m^4+116 a^2 m^2+126 a^4 \right)}{288 R_{\mathrm{Round}}^6}-\frac{m^2 a^2 \left(623295 m^4+186760 a^2 m^2+28368 a^4\right)}{8640 R_{\mathrm{Round}}^7} + O \left(\frac{1}{R_{\mathrm{Round}}^8}\right)} \, .
		\end{aligned}	
	\end{equation}
	We list the following properties of the spheres:
	\begin{enumerate} 
		\item Both spheres agree on leading orders, namely 
		\begin{equation}
			R_{\mathrm{Rigid}}-R_{\mathrm{Round}}=O\left(\frac{1}{R_{\mathrm{Round}}^{2}}\right)=O\left(\frac{1}{r^{2}}\right) \, .
		\end{equation}
		The spheres approach the same surface in the asymptotically flat regime.
		\item In the limit $m=0$, the Kerr spacetime becomes flat Minkowski spacetime. In this case, we find
		\begin{equation}
			\left. R_{\mathrm{Round}}^2 \right|_{m=0}=\left. R_{\mathrm{Rigid}}^2 \right|_{m=0}=r^2+a^2 \sin^2 \theta+O\left(\frac{1}{r^{7}}\right) \, .
		\end{equation}
		The above outcome is a positive result of the following test. Kerr spacetime in the limit $m\to0$ becomes Minkowski spacetime. It fulfills the condition (\ref{Round_Rigid_condition}). In the Minkowskian case, round and rigid spheres may be the same surface.
	\end{enumerate}
	
	Another issue to discuss is the existence of a well-defined coordinate system. The problem with the singularity of the coordinate system realizes the subtleties of the choice of the surface on which we measure the energy, including the proper selection of coordinates. In general, there exist two cases that can be problematic:
	\begin{enumerate}
		\item The choice of surface.\\
		The condition for inner or exterior geometry does not ensure a topology of the chosen surface. It can be degenerated in many ways. For example, the surface may contain self-intersections or cone singularities.
		\item The choice of coordinates.\\
		It is a well-known problem of how the coordinate singularity can be generated.
	\end{enumerate}
	Both of the above topics lead to the issue of the existence of the solution of partial differential equations. Also, any points that may cause the topological problem must be checked. If we obtain the solution in the form of a series, the above analysis is connected with the examination of the convergence of the series. This analysis will be presented elsewhere.

	\subsection{Summary}
	
	The paper's main goal was to introduce the Hamiltonian dynamics for spacetimes with asymptotically Kerr end (section \ref{variational_Kijowski}). In particular, the specific choice of the gauge is presented in the chapter \ref{sec: Gauge_conditions}, which is based on multipole expansion. It leads to the essential formula for the Hamiltonian (\ref{Hamiltonian_multipoled_final}).
	
	The Hamiltonian dynamics were examined for two particular choices of surfaces: Round Spheres and Rigid Spheres. The spheres were described in the section \ref{ssec: Round_Rigid_definition}. The results are presented respectively in \ref{ssec:Results_Round} and \ref{ssec:Results_Rigid}. The outcomes for both surfaces are compared in \ref{ssec:Comparsion_Results}.
	The observations from the section \ref{ssec:Comparsion_Results} will be used in the further examinations of surfaces. We planned to investigate mixed-geometry spheres--surfaces defined by interior and exterior geometry conditions. It will be discussed in a separate paper.
	
	\setcounter{secnumdepth}{0} %% no numbering
	\subsection{Acknowledgments}
	The authors thank Jerzy Kijowski for many fruitful discussions, especially for the help with the analysis of the embedding of the rigid sphere in the reference frame\footnote{The analysis of embedding is presented in the paragraph which contains equation \eqref{reference_coords_transform}. }.
	\setcounter{secnumdepth}{2}
	\newpage
	\appendix
	\section{Appendix}
	
	\subsection{$(n+1)$-decomposition of metric \label{time_decomposition}}
	The coordinate $x^{\alpha}$ will be decomposed from $(n+1)$-dimensional metric. In this section, greek letters (except $\alpha$) denote coordinates on $(n+1)$-dimensional manifold. The coordinates on hypersurface $\{x^{\alpha}=\mathrm{const.}\}$ run $k,l,...,z$, except $s$.\\
	We introduce the sign $s$, which is defined as follows
	\begin{equation}
		s=\left\{\begin{array}{r c}
			1 & \quad \mathrm{if} \quad x^{\alpha} x^{\alpha} g_{\alpha \alpha}>0\\
			-1 &\quad \mathrm{if} \quad x^{\alpha} x^{\alpha} g_{\alpha \alpha}<0
		\end{array}  \right.
	\end{equation}
	There is not a summation over $\alpha$. The null coordinate is not investigated. By $\nu$ and $\nu^{k}$ we denote respectively the lapse function and the shift vector in the $n$-dimensional geometry $g_{kl}$
	\begin{eqnarray}
		\nu&=&\frac{1}{\sqrt{|g^{\alpha \alpha}|}}\\
		\nu^{l}&=&\tilde{g}^{lm}g_{\alpha m}
	\end{eqnarray}
	where $\tilde{g}^{lm}$ is the inverse of $n$-dimensional metric. The $(n+1)$-dimensional Lorentzian metric $g_{\mu \nu}$ can be decomposed as
	\begin{equation}
		g_{\mu \nu}=
		\left[\begin{array}{cc}
			s \nu^2 +\nu^{m} \nu_{m}&\nu_{l} \\
			\nu_{k}&g_{kl}
		\end{array} \right] %\ ,
	\end{equation}
	and\\
	\begin{equation}
		g^{\mu \nu}=
		\left[\begin{array}{cc}
			\frac{s}{\nu^2}&- s \frac{\nu^{l}}{\nu^2} \\
			- s \frac{\nu_{k}}{\nu^2}&\tilde{g}^{kl}+ s \frac{\nu^{k} \nu^{l}}{\nu^{2}}
		\end{array} \right] %\ .
	\end{equation}
	\begin{equation}
		\sqrt{|\det g_{\mu \nu}|}=\nu \sqrt{|\det g_{k l}|}
	\end{equation}

	\subsection{Vector decomposition on the sphere: longitudinal and transversal part}
	The Hodge decomposition of the covector $\xi$ on a compact manifold is
	\begin{equation}
		\xi=\mathrm{d} \alpha+\delta\beta+h
	\end{equation}
	In case of the unit sphere with coordinates $(\theta,\varphi)$
	the harmonic one-form $h$ vanishes. %($\mathrm{d} h=0=\delta h$ implies $h=0$). 
	It is implied by the topology of the unit sphere (triviality of the corresponding cohomology class). Hence, we can always represent $\xi$ as follows
	\begin{equation}\label{xiab}
		\xi_A = \alpha_{,A} +\varepsilon_A{^B}\beta_{,B}
	\end{equation}
	where functions $\alpha$ and $\beta$ are defined up to a constant, but their gradients are unique. The $\alpha_{,A}=:\left({\! \hphantom{|}}_{|| \!} \xi \right)_{A}$ and $\varepsilon_A{^B}\beta_{,B}=:\left({\! \hphantom{|}}_{\perp \!} \xi \right)_{A}$ is called respectively the longitudinal part and the transversal part of the covector $\xi$ on the unit sphere. \\
	If the covector $\omega$ has axial symmetry, i.e., its components are $\varphi$-independent, then the longitudinal part of the covector has only the $\theta$-component, and its transversal part has only $\varphi$-component. More precisely
	\begin{eqnarray}
		\omega&=&\omega_{\theta}(\theta) \mathrm{d} \theta+\omega_{\varphi}(\theta) \mathrm{d} \varphi\\
		{\! \hphantom{|}}_{||} \omega&=& \omega_{\theta}(\theta) \mathrm{d} \theta \\
		{\! \hphantom{|}}_{\perp \!} \omega &=& \omega_{\varphi}(\theta) \mathrm{d} \varphi
	\end{eqnarray}
	
	\subsection{Multipole expansions on an arbitrary topological sphere \label{multipole_expansion}}
	It needs to be made clear how to perform a multipole decomposition on an arbitrary Riemannian two-dimensional manifold with a sphere's topology. We use the following method, which is based on equilibrated spherical coordinates. Our exposition closely follows \cite{Leski2013_Rigid_Spheres}.
	
	Let $\mathcal{S}$ be a differential two-manifold, diffeomorphic to the two-sphere $S^2\subset {\mathbb R}^3$ and equipped with a (sufficiently smooth) metric $g$. Coordinates $(\theta, \varphi) = (x^A)$, $A=1,2$, defined on a dense subset of $\mathcal{S} \setminus \ell$, where $\ell$ is topologically a line interval, will be called Conformally Spherical Coordinates if they have the same range of values as the standard spherical coordinates on $S^2 \subset {\mathbb R}^3$ and if the corresponding metric tensor $g_{AB}$ is conformally equivalent to the standard round metric on $S^2$, i.e. the following formula holds:
	\begin{equation}\label{conf0}
		g_{AB} = \Psi^2 \cdot \sigma_{AB} %\ ,
	\end{equation}
	where $\Psi^2$ is a (sufficiently smooth) function on $\mathcal{S}$ and
	\begin{equation}\label{eta}
		\sigma = \mathrm{d} \theta^2+ \sin^{2} \theta \mathrm \varphi^{2}
	\end{equation}
	Conformally spherical coordinates always exist. It is easy to see that there is
	always a six-parameter freedom in the choice of such coordinates.
	
	Given a system of conformally spherical coordinates on $\mathcal{S}$, consider the corresponding three functions:
	\begin{eqnarray}
		x &:=& \sin \theta \cos\varphi  \label{x}\\
		y &:=& \sin \theta \sin\varphi  \label{y}\\
		z &:=& \cos \theta \label{z}
	\end{eqnarray}
	We have, therefore, a mapping ${\mathbf F}\!: \, (0,\pi)\times(0,2\pi) \, \mapsto {\mathbb R}^3$, given by:
	\begin{equation}\label{dipolD}
		{\mathbf F}(\theta,\varphi)=
		\left(\begin{array}{c}
			F^1(\theta,\varphi)\\ F^2(\theta,\varphi)\\ F^3(\theta,\varphi)
		\end{array}\right) =
		\left(\begin{array}{c}
			x \\
			y \\
			z \end{array}\right)
	\end{equation}
	The following vector
	\begin{equation}\label{B}
		{\bf B} =
		\left( \begin{array}{c}
			< x >  \\   < y > \\ < z >
		\end{array} \right) \in {\mathbb R}^3
	\end{equation}
	whereby $< u >$ we denote the average (mean value) of the
	function $u$ on $\mathcal{S}$, i.e. the number
	\begin{equation}\label{average}
		<u> := \frac{\int_{\mathcal{S}} u \sqrt{\det g_{AB}}\ \mathrm{d}^2 x^{A}}
		{\int_{\mathcal{S}} \sqrt{\det g}\ \mathrm{d}^2 x}
	\end{equation}
	will be called a ``barycenter'' of the system $(\theta,\varphi )$ on $\mathcal{S}$.
	Of course, we have $\| {\bf B} \| \leq 1$, because of the H\"older inequality
	\[
	\| {\bf B} \|^2 \, =  \, <x>^2 + <y>^2 + <z>^2 \ \le \ <x^2> + <y^2> + <z^2> \,=\, 1
	\]
	\begin{df}
		Conformally spherical coordinate system $(\theta,\varphi)$ is called equilibrated, if its barycenter vanishes: ${\bf B} = 0\in {\mathbb R}^3$. \label{df_Equilibrated}
	\end{df}
	If there are two equilibrated spherical systems on $\mathcal{S}$, then they are related by a rotation. In the paper \cite{Leski2013_Rigid_Spheres}, the following theorems was proved:
	\begin{thm}
		Each metric tensor on $\mathcal{S}$ admits a unique (up to rotations) equilibrated spherical system.
	\end{thm}
	\begin{thm}\label{unique}
		Let $\mathcal{S}$ be a differential two-manifold, diffeomorphic to the two-sphere
		$S^2\subset {\mathbb R}^3$ and equipped with a metric $g$ of class $C^{k}$. For every pair ${\bf n}, {\bf m} \in S$, ${\bf n} \ne {\bf m}$, there is a unique equilibrated spherical system $(\theta,\varphi)$ of coordinates on $S$, such that $\theta$ vanishes at ${\bf n}$ and $\varphi$ vanishes at ${\bf m}$, and the metric components $g_{AB}$ are of the same class $C^{k}$.
	\end{thm}
	Given a manifold $S$ equipped with a metric tensor $g$, there is a	three-dimensional space of ``linear functions'' uniquely defined on $S$ as linear combinations of functions (\ref{x}--\ref{z}), calculated in any equilibrated spherical system of coordinates $(\theta,\varphi)$. We denote this space by
	${\cal M}^3$. By ${\cal M}^4$, we denote the space spanned by ${\cal M}^3$ and the constant functions on $S$. Linear functions (\ref{x}--\ref{z}) on ${S}$ are eigenfunctions of the Laplace operator\footnote{By $\Delta_{\sigma}$ we denote the usual Laplace operator for the unit-sphere metric (\ref{eta}).}  $\Delta_{\sigma}$, with the eigenvalue equal to $-2$, i.e.
	%\[
	$\Delta_{\sigma} X^i = -2 X^i$,
	%\]
	where we denote $x=X^1$, $y=X^2$, $z=X^3$. Let us denote by $\mbox{\rm d}\sigma:=\sin\theta\, {\rm d}\theta \, {\rm d}\varphi$ the measure associated with the metric $\sigma_{AB}$.
	\begin{df}
		\label{Def0} Let $f \in L^2(S,\mbox{\rm d}\sigma)$. The projection of $f$ onto the subspace of constant functions:
		\begin{equation}\label{mon}
			P_m(f) := \frac{1}{4\pi}\int_{{ S}}
			f \, \mbox{\rm d}\sigma
		\end{equation}
		will be called the {\em monopole part} of $f$, whereas the projection onto \\  ${\cal M}^3 = \lin \{ X^1,X^2,X^3 \}$:
		\begin{equation} \label{eq1c}
			P_d(f) := \sum_{i=1}^3 \left(X^i \frac{\int_{{ S}} X^i f
				\, \mbox{\rm d}\sigma}{\int_{{ S}} (X^i)^2 \, \mbox{\rm d}\sigma }\right)
		\end{equation}
		will be called the {\em dipole part} of $f$. In addition, we set
		\begin{equation}
			%%\begin{gather}
			\label{eq2c} {\cal M}^4 := \lin \{ 1 \}
			\oplus {\cal M}^3
			= \lin \{1,X^1, X^2 ,X^3  \} \, ,
		\end{equation}
		and  $P_{md}(f):= P_{m}(f) + P_{d}(f) \in {\cal M}^4$
		denotes the {\em mono-dipole part} of  $f$.
	\end{df}
	The above structure enables us to define the multipole
	decomposition of the functions defined on a topological sphere $S$
	in terms of the eigenspaces of the Laplace operator associated with
	the metric $\sigma_{AB}$. If $h$ is a function on $S$, then by
	$h^{\mathbf{m}}:= P_m(h)$ we denote its monopole (constant) part, by
	$h^{\mathbf{d}}:=P_d(h)$ --- the dipole part (projection to the eigenspace of the Laplacian
	with eigenvalue $-2$). By $h^{\mathbf{w}} := (I-P_{md})(h)= h -h^{\mathbf{m}}- h^{\mathbf{d}}$ we
	denote the ``wave'', or mono-dipole-free, part of $h$.\\
	On the other hand, the multipoles fulfill the Helmholtz equation
	\begin{equation}
		\left[\triangle_{\sigma} +l(l+1)\right] Y_l(n) = 0 \, , \quad n\in S^2
	\end{equation}
	By $l$, we denote the multipole index. The monopole is the solution of the Helmholtz equation for $l=0$, the dipole for $l=1$, etc. In the section \ref{Table_multipoles}, we present a table of the first few, $\varphi$ - independent multipoles.
	\paragraph{Generalization of the multipole decomposition for covector fields on $(S^{2},\sigma_{AB})$}
	Let us consider the following diagram:
	\begin{equation}
		\begin{array}{ccccc}
			V^0_{k+1}\oplus V^0_{k+1} & \stackrel{i_{01}}{\longrightarrow} & V^1_{k} & \stackrel{i_{10}}{\longrightarrow} & V^0_{k-1} \oplus V^0_{k-1} \\
			\Big\downarrow\vcenter{\rlap{$\scriptstyle Fl$}} &  &
			\Big\downarrow\vcenter{\rlap{$\hat{} $}} &  &
			\Big\downarrow\vcenter{\rlap{$\scriptstyle Fl$}}  \\
			V^0_{k+1}\oplus V^0_{k+1} & \stackrel{i_{01}}{\longrightarrow} & V^1_{k} & \stackrel{i_{10}}{\longrightarrow} & V^0_{k-1} \oplus V^0_{k-1}
		\end{array} \nonumber
	\end{equation}
	The mappings are defined as
	\begin{eqnarray}
		i_{01}(f,g)&=&f_{\wr\wr A}+\varepsilon_A{^B}g_{\wr\wr B} \label{multipole_cov} \\
		i_{10}(v)&=&\left( v^A_{\wr\wr A}, \varepsilon^{AB}v_{A\wr\wr B} \right)\\
		Fl(f,g)&=&(g,f)\\
		{\hat v}_A&=&\varepsilon_A{^B}v_{B}
	\end{eqnarray}
	where $f,g \in V^0_{m}$; $v \in V^1_{m}$ and by "$ \wr\wr $" we denote the covariant derivative associated with the metric $\sigma_{AB} $. The spaces are the following
	\begin{itemize}
		\item[$V^0_{k+1} $]-- scalars on $S^2$ belonging to H\"older space $C^{k+1,\alpha}$
		\item[$V^1_{k}$]-- covectors on $S^2$ belonging to H\"older space $C^{k,\alpha}$
	\end{itemize}
	Denote by $\triangle_{\sigma}$ the Laplace operator on $S^2$  and by $SH^l$ the space of
	spherical harmonics of degree $l$, ($f\in SH^l \Longleftrightarrow
	{\triangle_{\sigma}}f= -l(l+1)f$).
	The following equality
	\begin{equation}
		i_{10}\circ i_{01} = {\triangle_{\sigma}}
	\end{equation}
	shows that if we restrict ourselves to the spaces $\overline V^0:=V^0
	\ominus SH^0 = (I-P_{m})V^0$
	(${\triangle_{\sigma}}\overline V^0 =\overline V^0$) and
	$\overline V^1=V^1\ominus[i_{01}(SH^1)]$
	($({\triangle_{\sigma}}+I)\overline V^1=\overline V^1$), then, all the mappings in the above diagram become isomorphisms.
	Technically, the formula (\ref{multipole_cov}) describes the relation between the scalar multipoles and the covector multipoles, which possess axial symmetry. Let $X,Y \in SH^1$ and $X,Y$ are axially symmetric: $X=X(\theta), \, Y=Y(\theta)$ then
	\begin{eqnarray}
		[i_{01}(X,Y)]^{C}&=&g^{C A} X_{\wr\wr A}+\varepsilon{^{C B}}Y_{\wr\wr B}
	\end{eqnarray}
	If we put $X=Y=Y_{l}$, where $Y_{l}$ is a representative of a scalar multipole basis for given $l$, we receive the vector multipoles $L_{l}$
	\begin{equation}
		[L_{l}]^C=g^{C A} (Y_{l})_{\wr\wr A}+\varepsilon{^{C B}}(Y_{l})_{\wr\wr B}
	\end{equation}
	The first few representatives of axially symmetric vector multipoles are presented in section \ref{Table_multipoles}.
	\subsection{Table of multipoles \label{Table_multipoles}}
	The first few representatives of unnormalized, axially symmetric, scalar multipoles are
	\begin{eqnarray}
		Y_{0}(\theta)&=&1\\
		Y_{I}(\theta)&=& \cos \theta\\
		Y_{II}(\theta)&=&\left(3 \cos^{2} \theta-1\right)\\
		Y_{III}(\theta)&=&\left(5 \cos^{3} \theta-3 \cos \theta \right)\\
		Y_{IV}(\theta)&=&\left(35 \cos^{4} \theta-30 \cos^{2} \theta+3 \right)\\
		Y_{V}(\theta)&=&\left(63 \cos^{5} \theta-70 \cos^{3} \theta+15 \cos \theta \right)\\	
		Y_{VI}(\theta)&=&\left(231 \cos^{6} \theta-315 \cos^{4} \theta+105 \cos^{2} \theta -5 \right)\\
		Y_{VII}(\theta)&=&\left(429 \cos^{7} \theta -693 \cos^{5} \theta+315 \cos^{3} \theta-35 \cos^{} \theta\right)
	\end{eqnarray}
	We briefly remind the relations between representatives of the base of the scalar multipoles and the vector multipoles
	\begin{equation}
		\lambda \varepsilon^{\varphi \theta}= \sin \theta \varepsilon^{\varphi \theta}=-1 \qquad L^{\theta}_{l}=g^{\theta \theta} (Y_{l})_{,\theta} \qquad L^{\varphi}_{l}=\varepsilon^{\varphi \theta}(Y_{l})_{,\theta}
	\end{equation}
	where $l$ is the multipole index. In the paper, we are interested only in the $\varphi$ component of the multipole vector. The first few $\varphi$-components of the vector multipoles are the following
	\begin{eqnarray}
		L_{0}^{\varphi}&=& -\!\!\!-\!\!\!-\!\!\!-\\
		L_{I}^{\varphi}&=&1\\
		L_{II}^{\varphi}&=&\cos \theta\\
		L_{III}^{\varphi}&=&5 \cos^{2} \theta-1\\
		L_{IV}^{\varphi}&=&7 \cos^{3} \theta-3 \cos^{} \theta\\
		L^{\varphi}_{V}&=&21 \cos^{4} \theta-14 \cos^{2} \theta+1\\
		L^{\varphi}_{VI}&=&33 \cos^{5} \theta-30 \cos^{3} \theta+5 \cos^{} \theta\\
		L^{\varphi}_{VII}&=&429 \cos^{6} \theta-495 \cos^{4} \theta+135 \cos^{2} \theta-5
	\end{eqnarray}
	 
	\subsection{Rigid spheres in four dimensions}
	\label{sec:RigidFourDimAppendix}

	In section \ref{ssec: Round_Rigid_definition}, the concept of ``rigidity'' of spheres in generic Riemannian three-manifolds has been roughly discussed\footnote{Detailed discussion of the issue can be found in \cite{Leski2013_Rigid_Spheres}.}. This appendix presents a brief analysis of rigid spheres in four-dimensional Lorentzian spacetimes. See \cite{VietnamHPGJJJK} for the exhaustive considerations of the issue and the proof of the local existence of rigid spheres.
	
	\noindent

	\subsubsection{Preliminaries}
	
	In order to define a rigid sphere, some notation and geometrical objects need to be defined. Any choice of local coordinates $(x^0, \ldots ,x^3)$ on a subset of $\mathcal{M}$ induces the
	natural basis
	\[
	{\bf e}_\alpha := \frac \partial{\partial x^\alpha},
	\qquad  \alpha = 0, \dots 3
	\]
	in fibers of the tangent bundle $T \mathcal{M}$. Moreover, let ${\cal S} \subset \textbf{}$ be a {\em topological sphere}, i.e. a two-dimensional submanifold diffeomorphic to the {\em standard unit sphere} $S^2 \subset {\mathbb R}^3$. The diffeomorphism in question  $F: S^2 \rightarrow M$, it is locally described by four functions
	\begin{equation} \label{eq1a}
		F: (u^2,u^3) \mapsto x=(x^{\alpha} ) \in M \ , \
		x^{\alpha} = x^{\alpha} (u^2,u^3) \, ,
	\end{equation}
	where $(u^A)=(u^2,u^3)=(\theta, \phi)$ are standard spherical coordinates on $S^2$. Using parametrization (\ref{eq1a}), we may take vectors
	\begin{equation}
		\label{eq4a}
		{\bf b}_A = \partial_A x := \frac{\partial x}{\partial u^A } =
		\frac{\partial x^\alpha}{\partial u^A } {\bf e}_\alpha
	\end{equation}
	as a basis in each tangent space $T_x {\cal S}$. The matrices $b^{\alpha}_{A}$ give all relations between tangent spaces.
	Consequently, the {\em induced metric} $s = F^*g$ on ${\cal S}$ is represented by the metric tensor
	\begin{equation} \label{eq4}
		s_{AB} = g ({\bf b}_A ,{\bf b}_B) =
		\frac {\partial x^{\alpha}}{\partial u^A} \
		g_{\alpha \beta} \
		\frac {\partial x^{\beta}}{\partial u^B}
		\ .
	\end{equation}
	Associated with metric $s=(s_{A B})$, the covariant derivative $
	\overset{s}{\nabla}_A $ and the Laplace-Beltrami operator
	$\overset{s}{\Delta}:= \overset{s}{\nabla}{^A}
	\overset{s}{\nabla}_A = s^{A B} \overset{s}{\nabla}_{A}
	\overset{s}{\nabla}_B$ on ${\cal S}$ are given.
	To avoid confusion, by $\overset{g}{\nabla}$ and $\overset{s}{\nabla}$, we mean the covariant derivatives associated with metrices $g$ and $s$, respectively.

	\subsubsection{Description of extrinsic geometry}
	
	Let ${\cal S} \subset \mathcal{M}$ be a topological sphere. Then the {\em extrinsic curvature tensor} ${\bf K}_{A B}$ of ${\cal S}$ is defined as the orthogonal (concerning ${\cal S}$) part of the covariant derivative of the tangent vector field ${\bf b}_B$ in the direction of another tangent vector ${\bf b}_A$:
	\begin{equation}
		\label{eq4c} {\bf K}_{A B} :=
		\left( \overset{g}{\nabla}_{{\bf b}_A}
		{\bf b}_B \right)^\perp \in T{\cal S}^\perp \ .
	\end{equation}
	Especially convenient expression for the components of ${\bf K}_{A B}$ is obtained if we choose coordinates $(x^\alpha)$ in a
	neighbourhood of ${\cal S}$ in such a way that the transversal coordinates $x^0, x^1$ are constant on ${\cal S}$. In order to distinguish the normal directions, we introduce Greek indices with a tilde such that
	\begin{equation}
		\partial_{x^{\widetilde{\alpha}}} \in T{\cal S}^\perp \subset T\mathcal{M} \, .
	\end{equation}
	Consequently, ${\bf e}_2, {\bf e}_3$ are tangent to ${\cal S}$ and $(x^A)=(u^A)$. 
	Such coordinate system will be called {\em adapted to} ${\cal
		S}$.
	
	Hence, ${\bf b}_A= {\bf e}_A$ and
	\begin{equation} \label{eq6}
		\overset{g}{\nabla}_{{\bf b}_A}
		{\bf b}_B = \Gamma^{\alpha}_{A B} {\bf e}_{\alpha} =
		\Gamma^{\widetilde{\beta}}_{A B} {\bf e}_{\widetilde{\beta}} + \Gamma^C_{A B} {\bf e}_C \ ,
	\end{equation}
	with the Christoffel
	symbols $\Gamma^{\alpha}_{\beta \gamma}$ of  metric $g.$
	Using the projection ``$\perp$'' we have
	\begin{equation}\label{b_a}
		{\bf b}_{\widetilde{\beta}} := {\bf e}_{\widetilde{\beta}}^\perp = {\bf e}_{\widetilde{\beta}} - n_{\widetilde{\beta}}^{\ C} {\bf e}_C \, , \quad
		\text{where} \quad   n_{\widetilde{\beta}}^{\ C} = g_{{\widetilde{\beta}}B}\, s^{BC} \, ,
	\end{equation}
	as basis of the normal bundle $T{\cal S}^\perp$ and we obtain the
	following formula for the components $K^a_{A B}$ of the vector
	${\bf K}_{A B}=K^{\widetilde{\beta}}_{A B} {\bf b}_{\widetilde{\beta}}$, valid in adapted coordinates:
	\begin{equation}\label{kAB}
		K^a_{A B} = \Gamma^a_{A B} \ .
	\end{equation}
	\begin{df}
		\label{Def1} The {\em extrinsic curvature vector} ${\bf k}$ of ${\cal S}$ is defined by
		\begin{equation}\label{k}
			{\bf k} =
			%    \tr {\bf K} =
			s^{AB} {\bf K}_{A B} \in T{\cal S}^\perp \ ,
		\end{equation}
		Moreover, we consider its {\em extrinsic torsion}  at the point $x$:
		\begin{equation} \label{eq6b}
			t_A :=  g\big({\bf m} , \overset{g}{\nabla}_A {\bf n} \big) \ , \quad
			{\bf t}(x)=t_A {\rm d}u^A \in T^*_x{\cal S}
		\end{equation}
		where
		\begin{equation}\label{m}
			{\bf n} := \frac {\bf k}{\| {\bf k} \| } \qquad \text{({\em first} orthonormal vector)},
		\end{equation}
		${\| {\bf k} \| } = \sqrt{k^a g_{ab} k^b}$ provided ${\bf k}$ is
		space-like ($\|{\bf k}\|^2 := g(${\bf k}$,${\bf k}$)>0$),
		and ${\bf m}$ denotes the {\em second} orthonormal vector, i.e. the unit vector orthogonal to both ${\bf k}$ and $S$.
	\end{df}

	\subsubsection{Rigid spheres}
	The geometric objects defined in the previous paragraph enable us to formulate a rigorous definition of the rigid sphere. We continue with imposing conditions on two geometric scalar objects of a topological sphere ${\cal S}$:
	\begin{enumerate}
		\item the length $k:=\|{\bf k}\| = \sqrt{k^a s_{ab} k^b} $ of the extrinsic
		curvature vector, and
		\item the ``divergence'' of the extrinsic torsion
		\begin{equation}\label{q-div-t}
			{\bf q} = \ddiv {\bf t}:=
			\partial_A\left(\sqrt{\det s}\
			s^{AB} \ t_B \right) \, .
		\end{equation}
	\end{enumerate}
	Note that the above quantity does not depend upon the choice
	of the metric on ${\cal S}$, within the same conformal class.
	It leads to that the induced metric $s$ in the above formula can be replaced by the round sphere metric $\sigma$. The quantity ${\bf q}$ is a scalar density which is converted into a scalar by
	%%, we divide it by the volume density $ \sqrt{\det \sigma}$ and define:
	\begin{equation}\label{scalarq}
		q := \frac {1}{\sqrt{\det \sigma}} \ {\bf q} \, ,\quad \text{i.e.}
		\quad q= \overset{\sigma}{\nabla}_A t^A \, .
	\end{equation}

	\begin{df}\label{rigid}
		A topological sphere ${\cal S} \subset M$ will be called a
		{\em rigid sphere} if it possesses an equilibrated spherical coordinate system, given by definition \ref{df_Equilibrated} in appendix \ref{multipole_expansion}, such that the curvature vector ${\bf k}$ of ${\cal S}$ is space-like and that its both scalars $k$ and $q$ contain only the mono-dipole
		part\footnote{Mono-dipole part can be defined with the help of the equilibrated coordinate system. See definition in appendix \ref{multipole_expansion}.}. It means that the two following equations are satisfied:
		\begin{align}
			k_w=(I-P_{md}) \, k &= 0 &\quad \text{\em  (curvature condition)}\, , \label{md-k}\\
			q_w=(I-P_{md}) \, q &= 0 &\quad \text{\em  (torsion condition)}\, . \label{md-q}
		\end{align}
	\end{df}
	Note that the torsion scalar $q$ fulfills the
	identity
	$
	P_m (q) = \int_{\cal S} q \, \mbox{\rm d}\mu =0 \ ,
	$
	as a consequence of the definition \eqref{scalarq}. This means
	that the operator $P_{md}$ in \eqref{md-q} can be
	replaced by $P_d$.
	
	Both conditions, \eqref{md-k} and \eqref{md-q}, are obeyed by the
	eight-parameter family of round spheres in Minkowski space. \eqref{md-k} and \eqref{md-q} always admit an
	eight-parameter family of solutions if the spacetime metric is
	similar to the flat Minkowski metric.
		\subsection{The Kijowski's variational formula in a general case \label{variational_Kijowski}}
		
		We present a sketch of the variational formula derived and discussed in \cite{Kijowski_var_form} and \cite{Chr_Jez_Kij_Hamiltonian_Kerr_de_Sitter}. Let us consider the Lagrangian for vacuum Einstein equations:
		\begin{equation}
			L=\frac1{16 \pi} \sqrt{|g|} R=\pi^{\mu \nu} R_{\mu \nu}
		\end{equation}
		where $R$,$R_{\mu \nu}$ or $R_{\mu\nu\lambda \kappa}$ represent curvature scalar, Ricci tensor, or Riemann tensor, respectively. It is convenient to introduce
		\begin{equation}
			\pi^{\mu \nu}=\frac1{16 \pi} \sqrt{|g|} g^{\mu \nu}
		\end{equation}
		Consider a one-parameter family of field configurations $\epsilon \mapsto \left(g_{\mu,\nu}(\epsilon), \Gamma^{\kappa}_{\mu \nu }(\epsilon)\right)$ and define a ``variation'' $\delta$ as the operation
		\begin{eqnarray}
			\delta:= \left. \frac{\mathrm{d}}{\mathrm{d} \epsilon} \right|_{\epsilon=0}
		\end{eqnarray}
		thus $\delta g=\left. \frac{\mathrm{d}g}{\mathrm{d} \epsilon}\right|_{\epsilon=0}$, etc. The variation of the Lagrangian leads to the following equality
		\begin{eqnarray}
			\delta L &=& R_{\mu \nu} \delta \pi^{\mu \nu} + \pi^{\mu \nu} \delta R_{\mu\nu} \nonumber \\
			&=& -\frac1{16 \pi} \mathcal{G}^{\mu \nu} \delta g_{\mu \nu} -\left( \nabla_{\kappa} {P_{\lambda}}^{\mu \nu \kappa} \right) \delta \Gamma^{\lambda}_{\mu \nu} \nonumber \\
			& &+\partial_{\kappa}\left({P_{\lambda}}^{\mu \nu \kappa} \delta \Gamma^{\lambda}_{\mu \nu} \right) \label{EL_eq}
		\end{eqnarray}
		where
		\begin{equation}
			{P_{\lambda}}^{\mu \nu \kappa}=\delta_{\lambda}^{\kappa} \pi^{\mu \nu} -\delta_{\lambda}^{(\mu}\pi^{\nu) \kappa} \label{P}
		\end{equation}
		In the identity (\ref{EL_eq}) the following components can be identified:
		\begin{equation}
			\frac{\delta L}{\delta g_{\mu \nu}}= -\frac1{16 \pi} \mathcal{G}^{\mu \nu}=0 \label{Einstein_eq}
		\end{equation}
		are vacuum Einstein equations,
		\begin{equation}
			\frac{\delta L}{\delta \Gamma^{\lambda}_{\mu \nu}}=-\left( \nabla_{\kappa} {P_{\lambda}}^{\mu \nu \kappa} \right)=0 \label{LC_eq}
		\end{equation}
		are equivalent to that the connection $ \Gamma$ is metric. The equation (\ref{EL_eq}) holds regardless of whether the metric $g_{\mu \nu}$ itself satisfies the vacuum field equations or the connection is a Levi-Civita connection. Our further results assume that the conditions (\ref{Einstein_eq}), (\ref{LC_eq}) are fulfilled.
		For convenience, we introduce
		\begin{equation}
			A^{\lambda}_{\mu \nu}:=\Gamma^{\lambda}_{\mu \nu}-\delta^{\lambda}_{(\mu}\Gamma^{\kappa}_{\nu) \kappa} \label{def_A}
		\end{equation}
		$A^{\lambda}_{\mu \nu}$ is only a useful combination of connection coefficients. It has no specific geometrical interpretation. The variation of the Lagrangian (\ref{EL_eq}) takes the form
		\begin{equation}
			\delta L=\partial_{\lambda}\left(\pi^{\mu \nu} \delta A^{\lambda}_{\mu \nu} \right) \label{delta_L_A}
		\end{equation}
		We have chosen a vector field
		\begin{equation}
			X=\partial_{0} \label{X_d_t}
		\end{equation}
		as a vector field associated with the Hamiltonian flow. It is a vector field related to an observer in spatial infinity.
		To derive the `` Hamiltonian type'' variational identity, we rewrite the Lie derivative of the connection in the direction of  $X$
		\begin{equation}
			\mathcal{L}_{X} \Gamma^{\lambda}_{\mu \nu}=\nabla_{\mu}\nabla_{\nu} X^{\lambda} -X^{\sigma} {R^{\lambda}}_{\nu\mu \sigma}
		\end{equation}
		In terms of $A^{\kappa}_{\mu \nu}$ we have
		\begin{eqnarray}
			\pi^{\mu \nu} \mathcal{L}_{X} A^{\kappa}_{\mu \nu} &=&{P_{\lambda}}^{\mu \nu \kappa} \mathcal{L}_{X} \Gamma^{\lambda}_{\mu \nu} \nonumber \\
			&=& \left(\delta_{\lambda}^{\kappa} \pi^{\mu \nu}-\delta_{\lambda}^{\mu}
			\pi^{\kappa \nu}\right) \left(\nabla_{\mu}\nabla_{\nu} X^{\lambda}-X^{\sigma}{R^{\lambda}}_{\nu\mu \sigma}\right) \nonumber \\
			&=&\frac{\sqrt{|g|}}{16 \pi}\left[\nabla_{\mu} \left(\nabla^{\mu}X^{\kappa}-\nabla^{\kappa} X^{\mu}\right)+2 {R^{\kappa}}_{\sigma} X^{\sigma}\right] \nonumber \\
			&=&\frac{1}{16 \pi} \left\{\partial_{\mu}\left[ \sqrt{|g|} \left(\nabla^{\mu}X^{\kappa}-\nabla^{\kappa} X^{\mu}\right)\right] +2 \sqrt{|g|} {R^{\kappa}}_{\sigma} X^{\sigma} \right\} \label{Lie_A}
		\end{eqnarray}
		Combining Eq. (\ref{delta_L_A}) with Eq. (\ref{Lie_A}) we obtain
		\begin{eqnarray}
			- \delta(\pi^{\mu \nu} \mathcal{L}_{X} A^{\lambda}_{\mu \nu}-X^{\lambda} L)&=&\frac{1}{16 \pi} \left\{\partial_{\mu}\left[  \delta \sqrt{|g|} \left(\nabla^{\mu}X^{\kappa}-\nabla^{\kappa} X^{\mu}\right)\right] \right. \nonumber \\
			& & +\delta {\Big (}\underbrace{2 \sqrt{|g|} {R^{\kappa}}_{\sigma} X^{\sigma} +16 \pi X^{\lambda} L}_{-32 \pi X^{\sigma}{\mathcal{G}^{\lambda}}_{\sigma}=0}{\Big )} \Big \} \label{Eq_XL}
		\end{eqnarray}
		On the other hand (\ref{Eq_XL}) can be rewritten as
		\begin{eqnarray}
			- \delta(\pi^{\mu \nu} \mathcal{L}_{X} A^{\lambda}_{\mu \nu}-X^{\lambda} L)&=&\left(\mathcal{L}_{X} \pi^{\mu \nu}\right) \delta A^{\lambda}_{\mu \nu}- \left(\mathcal{L}_{X}A^{\lambda}_{\mu \nu}\right) \delta \pi^{\mu \nu} \nonumber \\
			& &+\partial_{\sigma} \left(X^{\lambda} \pi^{\mu \nu} \delta A^{\sigma}_{\mu \nu}-X^{\sigma} \pi^{\mu \nu} \delta A^{\lambda}_{\mu \nu}\right) \label{Eq_XL_Lie}
		\end{eqnarray}
		The term $\left(\mathcal{L}_{X} \pi^{\mu \nu}\right) \delta A^{\lambda}_{\mu \nu}- \left(\mathcal{L}_{X}A^{\lambda}_{\mu \nu}\right) \delta \pi^{\mu \nu}$ in (\ref{Eq_XL_Lie}) is the divergence of bivector density. Making use of (\ref{Eq_XL}) and (\ref{Eq_XL_Lie}) we obtain
		\begin{eqnarray}
			\left(\mathcal{L}_{X} \pi^{\mu \nu}\right) \delta A^{\lambda}_{\mu \nu}- \left(\mathcal{L}_{X}A^{\lambda}_{\mu \nu}\right) \delta \pi^{\mu \nu}&=&\partial_{\sigma} \left[\pi^{\mu\nu}(X^{\lambda} \delta A^{\sigma}_{\mu \nu}-X^{\sigma} \delta A^{\lambda}_{\mu \nu})\right] \nonumber \\ & &+\frac{1}{16 \pi}\partial_{\mu}\left[ \delta \sqrt{|g|} \left(\nabla^{\mu}X^{\lambda}-\nabla^{\lambda} X^{\mu}\right)\right] \label{div_biv_den}
		\end{eqnarray}
		We integrate (\ref{div_biv_den}) over $\mathcal{V}$. $\mathcal{V}$ is a subset of surface $\left\{x^{0}=\mathrm{const.}\right\}$. The boundary of $\mathcal{V}$ is a closed surface $\left\{x^{0}=\mathrm{const.}, x^{r}=\mathrm{const.}\right\}$. There is no summation convention over $0$ and over $r$.
		\begin{eqnarray}
			\int_{\mathcal{V}} \left[\left(\mathcal{L}_{X}A^{0}_{\mu \nu}\right) \delta \pi^{\mu \nu}-\left(\mathcal{L}_{X} \pi^{\mu \nu}\right) \delta A^{0}_{\mu \nu}\right] \mathrm{d}\Sigma_{0}&=&\frac{1}{16 \pi}\int_{\partial \mathcal{V}} \delta \sqrt{|g|} \left(\nabla^{r}X^{0}-\nabla^{0} X^{r}\right) \mathrm{d} \mathcal{S}_{0r} \nonumber \\
			& &+ \int_{\partial \mathcal{V}} \left[\pi^{\mu\nu}(X^{0} \delta A^{r}_{\mu \nu}-X^{r} \delta A^{0}_{\mu \nu})\right] \mathrm{d} \mathcal{S}_{0r} \nonumber \\
			&=&\frac{1}{16 \pi}\int_{\partial \mathcal{V}} \delta \sqrt{|g|} \left(\nabla^{r}X^{0}-\nabla^{0} X^{r}\right) \mathrm{d} \mathcal{S}_{0r} \nonumber \\
			& &+\int_{\partial \mathcal{V}} X^{0} \pi^{\mu\nu} \delta A^{r}_{\mu \nu} \mathrm{d}\mathcal{S}_{0r} \label{int_V}
		\end{eqnarray}
		where $\mathrm{d}\Sigma_{0}=\partial_{0}\lrcorner \Vol (g_{\mu \nu})$ and $\mathrm{d}\mathcal{S}_{0r}=\partial_{0} \wedge \partial_{r} \lrcorner \Vol (g_{\mu \nu})$. We turn our attention to the Komar-type boundary terms in (\ref{int_V})
		\begin{equation}
			\frac{1}{16 \pi}\delta \left[ \sqrt{|g|} \left(\nabla^{r}X^{0}-\nabla^{0} X^{r}\right)\right]=X^{0} \delta \left(\pi^{r \mu} \Gamma^{0}_{\mu 0}-\pi^{0 \mu} \Gamma^{r}_{\mu 0}\right)
		\end{equation}
		It holds that
		\begin{equation}
			\pi^{r \mu} \Gamma^{0}_{\mu 0}-\pi^{0 \mu} \Gamma^{r}_{\mu 0}=\frac{1}{8 \pi} \sqrt{|\det g_{cd}|} \tilde{g}^{0a}S_{0a}-\pi^{rr} \partial_{0} \left(\frac{\pi^{r 0}}{\pi^{rr}}\right) \label{Komar_Gamma}
		\end{equation}
		where $S_{0a}$ is defined in (\ref{extr_curv}). To continue, we transport connection coefficients under diffeomorphism $\Psi$. It has been shown in \cite{Kijowski_var_form}, for variations such that the image of $\mathcal{V}$ remains fixed (equivalently, $\delta \Psi^{0}=0$), that we have
		\begin{equation}
			\pi^{\mu \nu} \delta A^{0}_{\mu \nu}|_{\delta \Psi^{0}=0}=-\frac{1}{16 \pi} g_{kl} \delta P^{kl}+\partial_{k} \left[\pi^{00}\delta\left(\frac{\pi^{0k}}{\pi^{00}}\right)\right] \label{kij_var_id}
		\end{equation}
		The definition of external curvature of $\Sigma$ embedded in $\mathcal{M}$ in terms of $A^{0}_{mn}$ can be expressed as
		\begin{equation}
			K_{mn}:=-\frac1{\sqrt{|g^{00}|}} \Gamma^{0}_{mn}=-\frac1{\sqrt{|g^{00}|}} A^{0}_{mn}
		\end{equation}
		and a usual definition of ADM momentum
		\begin{equation}
			P^{kl}:=\sqrt{|\det g_{mn}|}\left(K\tilde{g}^{kl}-K^{kl}\right)
		\end{equation}
		We use (\ref{kij_var_id}) as follows: Consider any differentiable family of fields $s \mapsto (\pi(s),A(s))$. We note the identity
		\begin{equation}
			\partial_{0} A^{0}_{\mu \nu} \delta \pi^{\mu \nu}-\partial_{0} \pi^{\mu \nu}  \delta A^{0}_{\mu \nu}=-\partial_{0} (\pi^{\mu \nu} \delta A^{0}_{\mu \nu})+\delta (\pi^{\mu \nu} \partial_{0} A^{0}_{\mu \nu}) \label{identity_dt_A}
		\end{equation}
		Equation (\ref{kij_var_id}) implies
		\begin{equation}
			\partial_{0}\left( \pi^{\mu \nu} \delta A^{0}_{\mu \nu} \right)=\partial_{0} \left\{-\frac{1}{16 \pi} g_{kl} \delta P^{kl}+\partial_{k} \left[\pi^{00}\delta\left(\frac{\pi^{0k}}{\pi^{00}}\right)\right]\right\} \label{dt_kij_var_id}
		\end{equation}
		\begin{equation}
			\delta \left( \pi^{\mu \nu} \partial_{0} A^{0}_{\mu \nu} \right)=\delta \left\{-\frac{1}{16 \pi} g_{kl} \partial_{0} P^{kl}+\partial_{k} \left[\pi^{00} \partial_{0} \left(\frac{\pi^{0k}}{\pi^{00}}\right)\right]\right\} \label{delta_kij_var_id}
		\end{equation}
		Inserting the above (\ref{dt_kij_var_id}) and (\ref{delta_kij_var_id}) into the right hand side of (\ref{identity_dt_A}), we obtain
		\begin{eqnarray}
			\partial_{0} A^{0}_{\mu \nu} \delta \pi^{\mu \nu}-\partial_{0} \pi^{\mu \nu}  \delta A^{0}_{\mu \nu}&=&\frac{1}{16 \pi} \left[\partial_{0} g_{kl} \delta P^{kl}-\partial_{0} P^{kl} \delta g_{kl}\right] \nonumber \\ & &-\partial_{k} \left[\partial_{0} \pi^{00} \delta \left(\frac{\pi^{k 0}}{\pi^{00}}\right)-\partial_{0}   \left(\frac{\pi^{k 0}}{\pi^{00}}\right) \delta\pi^{00}\right] \label{var_id}
		\end{eqnarray}
		Using equations (\ref{Komar_Gamma}) and (\ref{var_id}), the equation (\ref{int_V}) takes the following form
		\begin{eqnarray}
			0&=&\frac{1}{16 \pi} \int_{\mathcal{V}} X^{0} \left[\partial_{0} g_{kl} \delta P^{kl}-\partial_{0} P^{kl} \delta g_{kl}\right] \mathrm{d}\Sigma_{0} \nonumber \\
			& &-\int_{\mathcal{V}} X^{0} \partial_{k} \left[\partial_{0} \pi^{00} \delta \left(\frac{\pi^{k 0}}{\pi^{00}}\right)-\partial_{0}   \left(\frac{\pi^{k 0}}{\pi^{00}}\right) \delta\pi^{00}\right] \mathrm{d}\Sigma_{0} \nonumber \\
			& & -\frac{1}{8 \pi} \int_{\partial \mathcal{V}} X^{0} \delta \left(\sqrt{|\det g_{cd}|} \tilde{g}^{0a}S_{0a} \right) \mathrm{d}\mathcal{S}_{0r} +\int_{\partial \mathcal{V}} X^{0} \delta \left[ \pi^{rr} \partial_{0} \left(\frac{\pi^{r 0}}{\pi^{rr}}\right) \right] \mathrm{d}\mathcal{S}_{0r} \nonumber \\
			& &-\int_{\partial \mathcal{V}} X^{0} \pi^{\mu\nu} \delta A^{r}_{\mu \nu} \mathrm{d}\mathcal{S}_{0r}\label{int_var_id}
		\end{eqnarray}
		The second line in (\ref{int_var_id}) can be transformed into a boundary term. To analyze it, we start with the identity
		\begin{equation}
			\pi^{00} \delta \left(\frac{\pi^{r 0}}{\pi^{00}}\right)+\pi^{rr} \delta \left(\frac{\pi^{r 0}}{\pi^{rr}}\right)=2 \sqrt{|\pi^{00}\pi^{rr}|} \delta \frac{\pi^{r0}}{\sqrt{|\pi^{00}\pi^{rr}|}}
		\end{equation}
		Writing
		\begin{equation}
			q:=\frac{\pi^{r0}}{\sqrt{|\pi^{00}\pi^{rr}|}}=\frac{g^{r0}}{\sqrt{|g^{00} g^{rr}|}}
		\end{equation}
		and assuming that $ \sgn \pi^{00} \sgn \pi^{rr}=-1$, one finds
		\begin{equation}
			2 \sqrt{|\pi^{00} \pi^{rr}|}=\frac{2}{16 \pi} \sqrt{|g|} \sqrt{|g^{00} g^{rr}|}=\frac{1}{8 \pi} \frac{\lambda}{\sqrt{1+q^2}}
		\end{equation}
		In this notation, we have
		\begin{equation}
			\pi^{00} \delta \left(\frac{\pi^{r 0}}{\pi^{00}}\right)+\pi^{rr} \delta \left(\frac{\pi^{r 0}}{\pi^{rr}}\right)=\frac{\lambda}{8 \pi} \frac{\delta q}{\sqrt{1+q^2}}=\frac{\lambda}{8 \pi} \delta \alpha \label{pi_to_alpha}
		\end{equation}
		where $\alpha:=\mathrm{\arsinh}(q)$ was defined in (\ref{alpha}). This and the calculation identical to the one leading to (\ref{var_id}) imply that the second line in (\ref{int_var_id}) takes the form
		\begin{equation}
			\int_{\partial \mathcal{V}} X^{0} \left[\partial_{0} \pi^{rr}\delta \left(\frac{\pi^{0r}}{\pi^{rr}}\right)- \partial_{0}\left(\frac{\pi^{0r}}{\pi^{rr}}\right) \delta \pi^{rr}+\frac{1}{8 \pi}\left(\partial_{0} \alpha \delta \lambda-\partial_{0} \lambda \delta \alpha \right) \right] \mathrm{d} \mathcal{S}_{0r} \label{alpha_boundary_term}
		\end{equation}
		For Kerr spacetime, the vector $\Psi_{\ast} \partial_{0}$ is tangent to the tube. It corresponds to $\alpha=0$. Eqs (\ref{int_var_id}) and (\ref{alpha_boundary_term}) give the intermediate result
		\begin{eqnarray}
			0&=&\frac{1}{16 \pi} \int_{\mathcal{V}} X^{0} \left[\partial_{0} g_{kl} \delta P^{kl}-\partial_{0} P^{kl} \delta g_{kl}\right] \mathrm{d}\Sigma_{0} \nonumber \\
			& &+\frac{1}{8 \pi} \int_{\partial \mathcal{V}} X^{0} \left(\partial_{0}\alpha \delta \lambda-\partial_{0}\lambda \delta \alpha\right) \mathrm{d} \mathcal{S}_{0r} \nonumber \\
			& &+\int_{\partial \mathcal{V}} X^{0} \left[\partial_{0} \pi^{rr}\delta \left(\frac{\pi^{0r}}{\pi^{rr}}\right)- \partial_{0}\left(\frac{\pi^{0r}}{\pi^{rr}}\right) \delta \pi^{rr} \right] \mathrm{d} \mathcal{S}_{0r} \nonumber \\
			& & -\frac{1}{8 \pi} \int_{\partial \mathcal{V}} X^{0} \delta \left(\sqrt{|\det g_{cd}|} \tilde{g}^{0a}S_{0a} \right) \mathrm{d}\mathcal{S}_{0r} +\int_{\partial \mathcal{V}} X^{0} \delta \left[ \pi^{rr} \partial_{0} \left(\frac{\pi^{r 0}}{\pi^{rr}}\right) \right] \mathrm{d}\mathcal{S}_{0r} \nonumber \\
			& &-\int_{\partial \mathcal{V}} X^{0} \pi^{\mu\nu} \delta A^{r}_{\mu \nu} \mathrm{d}\mathcal{S}_{0r} \label{intermediate_result}
		\end{eqnarray}
		Let us change the role of $x^{r}$ and $x^{0}$. There exist the following relations
		\begin{equation}
			A^{0}_{00}=\frac{1}{\pi^{00}} \left(\partial_{k} \pi^{0k}+A^{0}_{kl} \pi^{kl}\right) \label{A_000}
		\end{equation}
		\begin{equation}
			A^{0}_{0k}=-\frac{1}{2 \pi^{00}} \left(\partial_{k} \pi^{00}+2 A^{0}_{kl} \pi^{0l}\right) \label{A_00k}
		\end{equation}
		\begin{equation}
			A^{r}_{rr}=\frac{1}{\pi^{rr}} \left(\partial_{k} \pi^{rk}+A^{r}_{kl} \pi^{kl}\right) \label{A_nnn}
		\end{equation}
		\begin{equation}
			A^{r}_{rk}=-\frac{1}{2 \pi^{rr}} \left(\partial_{k} \pi^{rr}+2 A^{r}_{kl} \pi^{rl}\right) \label{A_nnk}
		\end{equation}
		Identities (\ref{A_000}) and (\ref{A_00k}) become constraints on the boundary of the world tube $\partial \Omega$, see Eqs (\ref{A_nnn}) and (\ref{A_nnk}). They imply
		\begin{eqnarray}
			\pi^{\mu \nu} \delta A^{r}_{\mu \nu}&=&\pi^{a b} \delta A^{r}_{a b}+2 \pi^{r a} \delta A^{r}_{r a}+\pi^{rr} \delta A^{r}_{r r} \nonumber \\
			&=&-\frac{1}{16 \pi} g_{ab}\delta Q^{ab}+\partial_{a}\left[\pi^{rr} \delta \left(\frac{\pi^{ra}}{\pi^{rr}}\right)\right] \label{pi_del_A}
		\end{eqnarray}
		where $Q^{ab}$ have been defined respectively in (\ref{ADM_cpart}). Equation (\ref{pi_del_A}) gives
		\begin{eqnarray}
			\int_{\partial_{\mathcal{V}}} X^{0} \pi^{\mu \nu} \delta A^{r}_{\mu \nu} \mathrm{d} \mathcal{S}_{0r}&=&-\frac{1}{16 \pi} \int_{\partial_{\mathcal{V}}} X^{0} g_{ab}\delta Q^{ab} \mathrm{d} \mathcal{S}_{0r}+ \int_{\partial_{\mathcal{V}}} X^{0}\partial_{a}\left[\pi^{rr} \delta \left(\frac{\pi^{ra}}{\pi^{rr}}\right)\right]\mathrm{d} \mathcal{S}_{0r} \nonumber \\
			&=&-\frac{1}{16 \pi} \int_{\partial_{\mathcal{V}}} X^{0} g_{ab}\delta Q^{ab}\mathrm{d} \mathcal{S}_{0r} + \int_{\partial_{\mathcal{V}}} X^{0}\partial_{A}\left[\pi^{rr} \delta \left(\frac{\pi^{r A}}{\pi^{rr}}\right)\right]\mathrm{d} \mathcal{S}_{0r} \nonumber \\
			& & +\int_{\partial_{\mathcal{V}}} X^{0} \delta \left[\pi^{rr} \partial_{0} \left(\frac{\pi^{0r}}{\pi^{rr}}\right) \right] \mathrm{d} \mathcal{S}_{0r}\nonumber \\
			& & + \int_{\partial_{\mathcal{V}}} X^{0} \left[\partial_{0} \pi^{rr} \delta \left(\frac{\pi^{0r}}{\pi^{rr}} \right)-\partial_{0} \left(\frac{\pi^{0r}}{\pi^{rr}}\right) \delta  \pi^{rr} \right] \mathrm{d} \mathcal{S}_{0r} \label{int_pi_del_A}
		\end{eqnarray}
		Combining Eqs (\ref{intermediate_result}) and (\ref{int_pi_del_A}) we obtain
		\begin{eqnarray}
			0&=&\frac{1}{16 \pi} \int_{\mathcal{V}} X^{0} \left[\partial_{0} g_{kl} \delta P^{kl}-\partial_{0} P^{kl} \delta g_{kl}\right] \mathrm{d}\Sigma_{0} \nonumber \\
			& &+\frac{1}{8 \pi} \int_{\partial \mathcal{V}} X^{0} \left(\partial_{0}\alpha \delta \lambda-\partial_{0}\lambda \delta \alpha\right) \mathrm{d} \mathcal{S}_{0r} \nonumber \\
			& & -\frac{1}{8 \pi} \int_{\partial \mathcal{V}} X^{0} \delta \left(\sqrt{|\det g_{cd}|} \tilde{g}^{0a}S_{0a} \right) \mathrm{d}\mathcal{S}_{0r} +\frac{1}{16 \pi} \int_{\partial_{\mathcal{V}}} X^{0} g_{ab}\delta Q^{ab}\mathrm{d} \mathcal{S}_{0r} \nonumber \\
			& &-\underbrace{\int_{\partial_{\mathcal{V}}} X^{0}\partial_{A}\left[\pi^{rr} \delta \left(\frac{\pi^{r A}}{\pi^{rr}}\right)\right]\mathrm{d} \mathcal{S}_{0r}}_{(\ast)} \label{almost_result}
		\end{eqnarray}
		With the help of Eqs from appendix \ref{time_decomposition}, one can show
		\begin{equation}
			(\ast)=-\int_{\partial_{\mathcal{V}}}\partial_{A} X^{0} N \sqrt{|\det g_{\mu \nu}|} g^{rr} \delta \left(\frac{g^{r A}}{g^{rr}}\right)\mathrm{d} \mathcal{S}_{0r}=0
		\end{equation}
		Eqs (\ref{bold_Q})--(\ref{bold_Q_2dim}) lead to
		\begin{equation}
			\frac{1}{8 \pi} \delta \left[\sqrt{|\det g_{cd}|} \left(\tilde{g}^{0a}S_{0a} \right) \right]-\frac{1}{16 \pi} g_{ab}\delta Q^{ab}=\frac{1}{16 \pi} \left(2 \nu \delta \textbf{Q}-2 \nu^{A}\delta \textbf{Q}_{A}+ \nu \textbf{Q}^{AB} \delta g_{AB} \right) \label{Q_id}
		\end{equation}
		We denote $\dot{ P^{kl}}=\partial_{0}  P^{kl}$ and put $X^{0}=1$. Eqs (\ref{almost_result}) and (\ref{Q_id}) enable one to reach the main result of this section
		\begin{eqnarray}
			0&=& \frac{1}{16 \pi} \int_{\mathcal{V}}\left(\dot{P}^{kl} \delta g_{kl}-\dot{g}_{kl} \delta P^{kl}\right) \mathrm{d}\Sigma_{0}+\frac{1}{8 \pi} \int_{\partial \mathcal{V}} \left(\dot{\lambda} \delta \alpha - \dot{\alpha} \delta \lambda\right) \mathrm{d} \mathcal{S}_{0r} \nonumber \\
			& &+\frac{1}{16 \pi} \int_{\partial \mathcal{V}} \left(2 \nu \delta \textbf{Q}-2 \nu^{A} \delta \textbf{Q}_{A}+ \nu \textbf{Q}^{AB} \delta g_{AB}\right) \mathrm{d} \mathcal{S}_{0r} \label{kij_var_form}
		\end{eqnarray}
		\subsection{Expected properties of quasi-local mass \label{ssec:QuasiMassProper}}
		A quasi-local mass should be a well-defined object for an arbitrary spacetime. When integrated over a two-dimensional spatial surface, it is expected to give a quantity that corresponds to our intuition about mass. Masses of well-examined spacetimes should cover with the result of quasi-local mass. One might adopt at least two strategies: trying to isolate a mathematically interesting object or finding a physically relevant one. \\
		Bartnik \cite{Bartnik1989new} has suggested a list of classical properties that should be satisfied by a good definition of quasi-local mass $m_{ql}(\mathcal{V})$. Suppose $\mathcal{V}$ is a domain in an asymptotically flat maximal slice $(\Sigma,g_{kl},K_{kl})$. The properties that one would like the quasi-local mass to satisfy are the following:
		\begin{enumerate}
			\item  \textit{Positivity\footnote{Positivity is particularly relevant in the context of the initial value problem in General Relativity, where the positivity of energy is an essential condition for the well-posedness of the evolution equations.}:} $m_{ql} \geq 0$.
			\item \textit{Vacuum distinguish:} $m_{ql} (\mathcal{V}) = 0$ if and only if $(\mathcal{V},g)$ is flat.
			\item \textit{Monotonicity:} $m_{ql}(\mathcal{V}) \leq m_{ql} (\tilde{\mathcal{V}})$ whenever $\mathcal{V} \subset \tilde{\mathcal{V}}$.
			\item \textit{Spherical mass:} $m_{ql}$ should agree with the spherical mass, like Misner--Sharp, on spherically symmetric balls or	annuli.
			\item \textit{ADM limit:} $m_{ql}$ should be asymptotic to the ADM mass.
		\end{enumerate}
		\bibliography{BibTeX_mass}

\providecommand{\bysame}{\leavevmode\hbox to3em{\hrulefill}\thinspace}
\providecommand{\MR}{\relax\ifhmode\unskip\space\fi MR }
\providecommand{\MRhref}[2]{%
  \href{http://www.ams.org/mathscinet-getitem?mr=#1}{#2}
}
\providecommand{\href}[2]{#2}
\begin{thebibliography}{10}

\bibitem{anderson2008boundary}
M.~Anderson, \emph{On boundary value problems for {E}instein metrics}, Geometry
  \& Topology \textbf{12} (2008), no.~4, 2009--2045.

\bibitem{Bartnik1989new}
R.~Bartnik, \emph{New definition of quasilocal mass}, Phys. Rev. Lett.
  \textbf{62} (1989), no.~20, 2346.

\bibitem{Bondi1962gravitational}
H.~Bondi, M.G.J. Van~der Burg, and A.W.K. Metzner, \emph{Gravitational waves in
  {G}eneral {R}elativity. {VII}. {W}aves from axi-symmetric isolated systems},
  Proc. Roy. Soc. London Ser. A, vol. 269, The Royal Society, 1962, pp.~21--52.

\bibitem{Cahill1970spherical}
M.~Cahill and G.~McVittie, \emph{Spherical symmetry and mass-energy in
  {G}eneral {R}elativity.}, J. Math. Phys. \textbf{11} (1970), no.~4,
  1382--1391, I. General Theory.

\bibitem{ChenWangYau2016}
P.-N. Chen, M.-T. Wang, and S-T. Yau, \emph{Quasi-local energy in presence of
  gravitational radiation}, Int. J. Mod. Phys. D \textbf{25} (2016), no.~13,
  1645001--.

\bibitem{Chen2011evaluating}
PN. Chen, MT. Wang, and ST. Yau, \emph{Evaluating quasilocal energy and solving
  optimal embedding equation at null infinity}, Comm. math. phys. \textbf{308}
  (2011), no.~3, 845--863.

\bibitem{Chrusciel2003hamiltonian_book}
P.~T. Chru{\'s}ciel, J.~Jezierski, and J.~Kijowski, \emph{Hamiltonian field
  theory in the radiating regime}, vol.~70, Springer Science \& Business Media,
  2003.

\bibitem{Chr_Jez_Kij_Hamiltonian_Kerr_de_Sitter}
\bysame, \emph{Hamiltonian dynamics in the space of asymptotically {K}err--de
  {S}itter spacetimes}, Phys. Rev. D \textbf{92} (2015), no.~8, 084030.

\bibitem{Chrusciel_lecture_notes}
P.T. Chru{\'s}ciel, \emph{Lectures on energy in {G}eneral {R}elativity
  {K}rak\'ow, march-april 2010}, Lecture Notes (2010).

\bibitem{dougan1991quasilocal}
A.J. Dougan and L.J. Mason, \emph{Quasilocal mass constructions with positive
  energy}, Phys. Rev. Lett. \textbf{67} (1991), no.~16, 2119.

\bibitem{DunajskiTod2021}
M.~Dunajski and P.~Tod, \emph{The {K}ijowski-{L}iu-{Y}au quasi-local mass of
  the {K}err black hole horizon}, Class. Quantum Grav. \textbf{38} (2021),
  no.~23, 235001--.

\bibitem{Geroch_mon_inv_mean_curv}
R.~Geroch, \emph{Energy extraction}, Ann. NY Academy of Sciences \textbf{224}
  (1973), 108--17.

\bibitem{geroch1973space}
R.~Geroch, A.~Held, and R.~Penrose, \emph{A space-time calculus based on pairs
  of null directions}, J. Math. Phys. \textbf{14} (1973), no.~7, 874--881.

\bibitem{VietnamHPGJJJK}
H.~P. Gittel, J.~Jezierski, and J.~Kijowski, \emph{On the existence of rigid
  spheres in four-dimensional spacetime manifolds}, Vietnam J. Math.
  \textbf{44} (2016), 231--249.

\bibitem{Leski2013_Rigid_Spheres}
H.~P. Gittel, J.~Jezierski, J.~Kijowski, and S.~{\L}{\k{e}}ski, \emph{Rigid
  spheres in {R}iemannian spaces}, Class. Quantum Grav. \textbf{30} (2013),
  no.~17, 175010.

\bibitem{Hawking_mass}
S.W. Hawking, \emph{Gravitational radiation in an expanding universe}, J. Math.
  Phys. \textbf{9} (1968), no.~4, 598--604.

\bibitem{Hernandez1966observer}
W.C. Hernandez~Jr and C.W. Misner, \emph{Observer time as a coordinate in
  relativistic spherical hydrodynamics}, Astrophys. J. \textbf{143} (1966),
  452.

\bibitem{Huang2010_mean_spheres}
L.~Huang, \emph{Foliations by stable spheres with constant mean curvature for
  isolated systems with general asymptotics}, Communications in Mathematical
  Physics \textbf{300} (2010), no.~2, 331--373.

\bibitem{Huisken2001inverse}
G.~Huisken and T.~Ilmanen, \emph{The inverse mean curvature flow and the
  {R}iemannian {P}enrose inequality}, J. Diff. Geom. \textbf{59} (2001), no.~3,
  353--437.

\bibitem{Jezierski1994stability}
J.~Jezierski, \emph{Stability of {R}eissner--{N}ordstr{\"o}m solution with
  respect to small perturbations of initial data}, Classical and Quantum
  Gravity \textbf{11} (1994), no.~4, 1055.

\bibitem{Jezierski1987_energy_positivity2}
J.~Jezierski and J.~Kijowski, \emph{Positivity of total energy in {G}eneral
  {R}elativity}, Physical Review D \textbf{36} (1987), no.~4, 1041.

\bibitem{Jezierski1987_energy_positivity1}
\bysame, \emph{A simple argument for positivity of gravitational energy},
  Proceedings of the XV International Conference on Differential Geometric
  Methods in Theoretical Physics (Clausthal, 1986), World Sci. Pub., Teaneck
  NJ, 1987, pp.~187--194.

\bibitem{jezierski1990_linear_hamiltonian}
\bysame, \emph{The localization of energy in gauge field theories and in linear
  gravitation}, General Relativity and Gravitation \textbf{22} (1990), no.~11,
  1283--1307.

\bibitem{Kijowski1986_energy_positivity}
J.~Kijowski, \emph{On positivity of gravitational energy}, Proc. of the IV
  Marcel Grossman Meeting on General Relativity (Rome, 1985), ed. R. Ruffini,
  North-Holland, Amsterdam, 1986, pp.~1681--1686.

\bibitem{Kijowski_var_form}
\bysame, \emph{A simple derivation of canonical structure and quasi-local
  {H}amiltonians in {G}eneral {R}elativity}, Gen. Relativ.Gravit. \textbf{29}
  (1997), no.~3, 307--343.

\bibitem{kijowski2002consistent}
\bysame, \emph{A consistent canonical approach to gravitational energy},
  Advances in General Relativity and Cosmology, Pitagora Bologna, Italy, 2002,
  pp.~129--145.

\bibitem{Kijowski_symp_framework}
J.~Kijowski and W.M. Tulczyjew, \emph{A symplectic framework for field
  theories}, Lect. Not. Phys. \textbf{107} (1979).

\bibitem{Komar_mass}
A.~Komar, \emph{Positive-definite energy density and global consequences for
  {G}eneral {R}elativity}, Phys. Rev. \textbf{129} (1963), no.~4, 1873--76.

\bibitem{Liu2003positivity}
CC.M. Liu and ST. Yau, \emph{Positivity of quasilocal mass}, Phys. Rev. Lett.
  \textbf{90} (2003), no.~23, 231102.

\bibitem{ludvigsen1983momentum}
M.~Ludvigsen and J.A.G. Vickers, \emph{Momentum, angular momentum and their
  quasi-local null surface extensions}, J. Phys. A \textbf{16} (1983), no.~6,
  1155.

\bibitem{Misner1964relativistic}
C.W. Misner and D.H. Sharp, \emph{Relativistic equations for adiabatic,
  spherically symmetric gravitational collapse}, Phys. Rev. \textbf{136}
  (1964), no.~2B, B571.

\bibitem{Neves_HG_mass_conv_spher_const_curv}
A.~Neves, \emph{Insufficient convergence of inverse mean curvature flow on
  asymptotically hyperbolic manifolds}, J. Diff. Geom. \textbf{84} (2010),
  no.~1, 191--229.

\bibitem{Nirenberg1953weyl}
L.~Nirenberg, \emph{The {W}eyl and {M}inkowski problems in differential
  geometry in the large}, Commun. Pure Appl. Math. \textbf{6} (1953), no.~3,
  337--394.

\bibitem{Murchadha2004comment}
N~. {\'O}~Murchadha, L.B. Szabados, and K.P. Tod, \emph{Comment on ''positivity
  of quasilocal mass''}, Phys. Rev. Lett. \textbf{92} (2004), no.~25, 259001.

\bibitem{Penrose_list}
R.~Penrose, \emph{Some unsolved problems in classical {G}eneral {R}elativity,
  seminar on {D}ifferential {G}eometry}, Ann. Math. Stud. \textbf{102} (1982),
  631--68.

\bibitem{Pogorelov1951izgibanie}
A.~V. Pogorelov, \emph{Izgibanie vypuklyh poverhnoste{\i}, {G}osudarstv},
  Izdat. Tehn.-Teor. Lit. (1951).

\bibitem{Sachs1962gravitational}
R.K. Sachs, \emph{Gravitational waves in {G}eneral {R}elativity. {VIII}.
  {W}aves in asymptotically flat space-time}, Proc. Roy. Soc. London Ser. A,
  vol. 270, The Royal Society, 1962, pp.~103--126.

\bibitem{Sauter_mon_con_null_hyp}
J.~Sauter, \emph{Foliations of null hypersurfaces and the {P}enrose
  inequality}, Ph.D. thesis, ETH, Zurich, Switzerland, 2008.

\bibitem{2018smojez_hopf}
T.~Smo{\l}ka and J.~Jezierski, \emph{Simple description of generalized
  electromagnetic and gravitational hopfions}, Class. Quant. Grav. \textbf{35}
  (2018), no.~24, 245010.

\bibitem{Szabados2009quasi}
L.B. Szabados, \emph{Quasi-local energy--momentum and angular momentum in
  {G}eneral {R}elativity}, Living Rev. Relativity \textbf{12} (2009), no.~4,
  90404.

\bibitem{Trautman1958}
A.~Trautman, \emph{Radiation and boundary conditions in the theory of
  gravitation}, Bull. Acad. Pol. Sci., S\'erie sci. math., astr. et phys.
  \textbf{VI} (1958), 407--12.

\bibitem{Trautman2002king}
\bysame, \emph{{K}ing`s {C}ollege lecture notes on {G}eneral {R}elativity,
  may-june 1958, mimeographed notes}, Rel. Grav \textbf{34} (2002), 721--762.

\bibitem{Wang2006generalization}
M.T. Wang and ST. Yau, \emph{A generalization of {L}iu--{Y}au's quasi-local
  mass}, arXiv:math/0602321 (2006).

\bibitem{Wang2009isometric}
\bysame, \emph{Isometric embeddings into the {M}inkowski space and new
  quasi-local mass}, Comm. math. phys. \textbf{288} (2009), no.~3, 919--942.

\bibitem{Wang2010limit}
\bysame, \emph{Limit of quasilocal mass at spatial infinity}, Comm. math. phys.
  \textbf{296} (2010), no.~1, 271--283.

\bibitem{York1974covariant}
J.~York~Jr, \emph{Covariant decompositions of symmetric tensors in the theory
  of gravitation}, Annales de l'IHP Physique th{\'e}orique, vol.~21, 1974,
  pp.~319--332.

\end{thebibliography}
\end{document}